\pdfoutput=1
\documentclass{article}

\RequirePackage{fix-cm}
\usepackage[english]{babel}
\usepackage[utf8]{inputenc}
\usepackage{amsmath}
\usepackage{amssymb}
\usepackage{graphicx}
\usepackage{pdfpages}
\usepackage{dsfont}
\usepackage{xcolor}
\usepackage{tikz}
\usetikzlibrary{shapes.geometric,calc,shadows.blur,decorations.pathmorphing}
\usepackage{subcaption}
\usepackage{geometry}
\usepackage{authblk}

\usepackage{ulem}

\newcommand{\beginAppendix}{
        \setcounter{table}{0}
        \renewcommand{\thetable}{A\arabic{table}}
        \setcounter{figure}{0}
        \renewcommand{\thefigure}{A\arabic{figure}}
     }

\title{NANOG/GATA6 Interactions Revisited: A Statistical Mechanics Approach towards Cell Fate Decisions}
\date{\today}

\begin{document}

\author[1]{Simon Schardt}
\author[1]{Sabine C. Fischer}
\affil[1]{\textit{Center for Computational and Theoretical Biology, Faculty of Biology, University of W\"urzburg, 97074 W\"urzburg, Germany}}

\maketitle

\begin{abstract}
In preimplantation mammalian embryos, the second cell fate decision introduces spatial patterns of embryonic and extra-embryonic precursor cells. The transcription factors NANOG and GATA6 are the earliest markers for the two cell types and interact between cells via the fibroblast growth factor signaling pathway. Computational models have been used to mimic the patterns and cell type proportions found in experimental studies. However, these models are always phenomenological in nature and lack a proper physical explanation. We derive a cell fate decision model motivated by the ideas of statistical mechanics. The model incorporates intra- and intercellular interactions of NANOG and GATA6. A detailed mathematical analysis on the resulting dynamical system is presented. We find that our model is capable of generating tissue wide spatial patterns of the two cell types. Its advantages are revealed in the simple physical and biological interpretation of the parameters and their interactions. In numerical simulations, we showcase the ability to replicate checkerboard patterns of different cell type proportions varying only a single parameter. The tight control of the system as well as the ease of use and the direct expandability to other signaling types provide solid reasons for the continued use of our model.
We are convinced that our approach presents an exciting perspective in relation to cell fate decisions. Moreover, the concepts are generalizable to questions regarding cell signaling beyond the mammalian embryo.
\end{abstract}

\section{Introduction}
In mammalian preimplantation development, cell differentiation plays a central role in the creation of blastocysts. In the mouse embryo, the first two cell fate decisions are conceptualized as two distinct events. In the first decision, cells separate to form the inner cell mass (ICM) and the trophoectoderm (TE). The second decision focuses on the pluripotent stem cells in the ICM. Here, cells differentiate into either embryonic precursor cells in the form of epiblast (Epi) cells or alternatively into extra-embryonic precursor cells in the form of primitive endoderm (PrE) cells \cite{Lanner2014,Perez2014,Frum2015,Chazaud2016}.

ICM organoids, a collection of mouse embryonic stem cells capable of organizing themselves into a sphere-like structure, show similarities to the in vivo system in terms of cell differentiation to Epi and PrE cells \cite{Mathew2019}. The large number of cells in a single ICM organoid is not only appealing to statistical analysis but also to generalized modeling approaches.

The first markers of Epi and PrE cell fates are the transcription factors NANOG and GATA6. The expression of NANOG plays a central role in the specification of Epi cells \cite{Mitsui2003}, whereas GATA6 is essential for PrE cells \cite{Schrode2014}. The expression of both transcription factors is controlled by a complex gene regulatory network (GRN). At the heart of the GRN resides the mutual inhibition of NANOG and GATA6 at the intracellular level. Cell-cell communication in the form of intercellular signaling allows for cells to influence neighboring cells. The fibroblast growth factor / extracellular signal-regulated kinase (FGF/ERK) pathway handles the task of communication, allowing for the formation of spatial patterns of the two different cell types. The impact of FGF4 on either cell fate has been investigated in experimental studies, showing the possibility to force cells to adopt either fate \cite{Nichols2009, Yamanaka2010}.

Computational models have proven capable of capturing the cell fate specification up to some extent, showing the possibility to create cell type proportions in a checkerboard pattern \cite{Bessonnard2014, Tosenberger2017}. Combined with few successive rules, relevant features of mammalian blastocysts have already been reconstructed in simulations \cite{Nissen2017, Saiz2020}. However, these models are always phenomenological in nature. Their heavy reliance on various applications of the Hill equation might introduce nonphysical behavior and neglects the characteristics of interactions between multiple constituents. A suitable physical description of the underlying mechanics as well as a rigorous mathematical analysis of the resulting equations are still pending.

We introduce a mathematical description of the antagonistic effect between NANOG and GATA6 using statistical mechanics \cite{Garcia2011}. We cover the general description of transcription factor binding up to the competitive or cooperative effect of the different interacting species. This leads to the description of binding probabilities, which enable us to set up a system of ordinary differential equations (ODEs) describing the concentration of NANOG and GATA6 in a physically meaningful way. These specifically derived binding probabilities allow us to distance ourselves from existing models \cite{Bessonnard2014, Tosenberger2017, Stanoev2021}. A detailed linear stability analysis leads to an elegant parameter restriction for homogeneous and heterogeneous steady states incorporating all of the model parameters. As a result, we get full control over the proportions of the two cell types. An extension in the functionality of the FGF4 signaling suffices to display the checkerboard pattern observed in the existing models. Finally, robustness of the model in terms of parameter changes and tissue sizes is showcased.

\section{Model Derivation}

\subsection{Transcription factor binding}

We consider the problem of transcription factor binding in a gene regulatory network (GRN) in terms of statistical mechanics \cite{Garcia2011}. Dividing our space into $\Omega$ different lattice sites, a number of transcription factors $A$ can rearrange in that space in
$$\frac{\Omega !}{A! (\Omega - A)!}$$
ways. Assuming there is only one binding site on the DNA for $A$ to bind to, then the number of different microstates in which $A$ is bound is
$$\frac{\Omega !}{(A-1)! (\Omega - A + 1)!}.$$
We assume there is an energetic difference in the bound and unbound state. Therefore, we introduce the energies for the unbound and bound state as $\varepsilon_{A,u}$ and $\varepsilon_{A,b}$, respectively. A state with no bound $A$ will then have an energy of
$$\varepsilon_1 = A\varepsilon_{A,u},$$
whereas for a bound state we get
$$\varepsilon_2 = (A-1)\varepsilon_{A,u} + \varepsilon_{A,b}.$$
In statistical mechanics, the partition function is given by the sum of all possible Boltzmann weights $e^{-\beta \varepsilon_i}$ over every microstate, i.e.
\begin{align*}
    Z_{total} &= \sum_{\text{microstates}} e^{-\beta\varepsilon_{\text{microstate}}} \\
    &= \frac{\Omega!}{A!(\Omega-A)!}e^{-\beta\varepsilon_1}  +\frac{\Omega!}{(A-1)!(\Omega-A+1)!}e^{-\beta\varepsilon_2} \\
    &= Z_1 + Z_2.
\end{align*}
The binding probability $p_A$ is then given by
$$p_A = \frac{Z_2}{Z_1 + Z_2}.$$
Up to this point, the procedure is very general in nature, i.e. find the number of microstates according to your GRN and define your partition function. The binding probabilities for any species is then found by dividing its part of the partition function by the total. Assuming $\Omega \ll A$, we can use the approximation $\frac{\Omega !}{(\Omega-A)!} \approx \Omega^A$. We divide numerator and denominator by $Z_1$ and define the energy difference $\Delta\varepsilon_A := \beta(\varepsilon_{A_b} - \varepsilon_{A_u})$.
\begin{equation}
    p_A = \frac{Z_2/Z_1}{1 + Z_2/Z_1} = \frac{\frac{A}{\Omega}e^{-\Delta\varepsilon_a}}{1 + \frac{A}{\Omega}e^{-\Delta\varepsilon_a}}.
\end{equation}
For simplicity, we replace the exponential expression with the following energy coefficient $\eta_a := e^{-\Delta\varepsilon_a}$ and use $a=A/\Omega$ to get the volume fractions. Here, $a=1$ would represent a fully occupied space, where $a=0$ resembles empty space. This leads to
\begin{equation}
    \label{eq: Hill equation}
    p_A = \frac{\eta_a a}{1 + \eta_a a}.
\end{equation}
This is the well-known Hill equation that is also commonly used in the same context \cite{Bessonnard2014,Tosenberger2017,Stanoev2021,Cang2021}. In the following we will refer to this as $p_A^{hill}$.

\subsection{Interactions}

The crucial parts in transcriptional regulation are the interactions between constituents. In the following, we consider two possibly interacting species $A$ and $B$. This results in a system, with the following microstates
\begin{enumerate}
    \item Neither $A$ nor $B$ are bound. \# of combinations = $\frac{\Omega!}{A!B!(\Omega-A-B)!}$
    \item $A$ is bound. \# of combinations = $\frac{\Omega!}{(A-1)!B!(\Omega-A-B+1)!}$
    \item $B$ is bound. \# of combinations = $\frac{\Omega!}{A!(B-1)!(\Omega-A-B+1)!}$
    \item $A$ and $B$ are bound. \# of combinations = $\frac{\Omega!}{(A-1)!(B-1)!(\Omega-A-B+2)!}$
\end{enumerate}

\begin{figure}[htbp]
\centering
\includegraphics[width=0.5\textwidth]{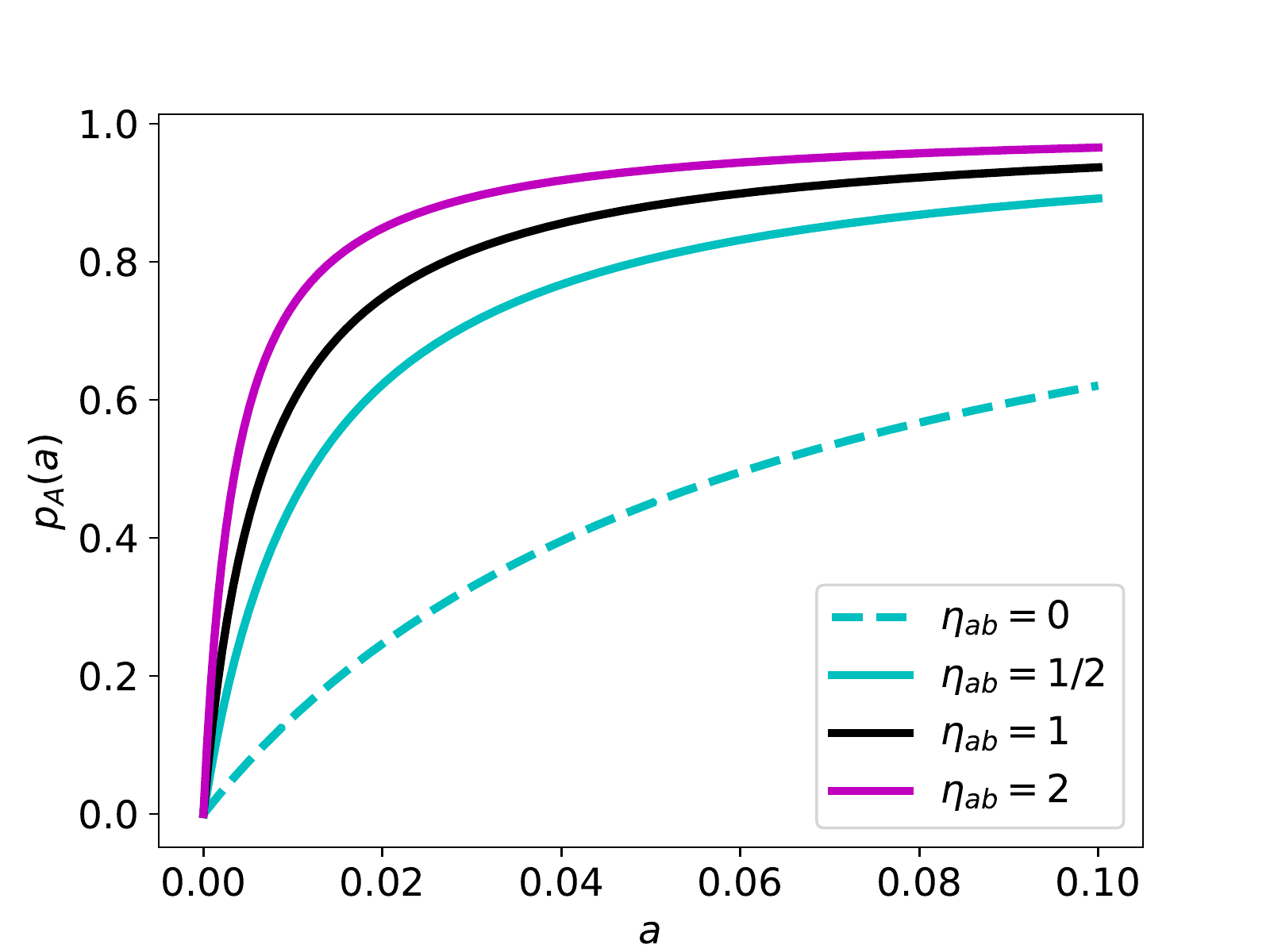}
\caption{Binding probabilities for different interaction coefficients. Plots were generated using a constant value of $b=0.02$ as well as $-\Delta\varepsilon_a = 5$ and $-\Delta\varepsilon_b = 6$. The black line represents the hill function \eqref{eq: interaction_hill}.}
\label{fig: probability interaction}
\end{figure}

The binding energy differences remain as before with an additional factor for the interaction $\eta_{ab} = e^{-\Delta\varepsilon_{ab}}$. The binding probabilities for $A$ and $B$ are then given by
\begin{align}
    p_A &= \frac{\eta_a a + \eta_a \eta_b \eta_{ab} ab}{1 +\eta_a a + \eta_b b + \eta_a \eta_b \eta_{ab} ab} \\
    p_B &= \frac{\eta_b b + \eta_a \eta_b \eta_{ab} ab}{1 +\eta_a a + \eta_b b + \eta_a \eta_b \eta_{ab} ab}.
\end{align}
The advantage or disadvantage given by the interaction energy difference now determines the nature of the interaction:

\begin{align*}
    &\eta_{ab} = 0 \quad \Longleftrightarrow \quad \Delta\varepsilon_{ab} = \infty &&\quad\text{complete inhibition / blocking},\\
    &\eta_{ab} < 1 \quad \Longleftrightarrow \quad \Delta\varepsilon_{ab} > 0 &&\quad\text{inhibition},\\
    &\eta_{ab} = 1 \quad \Longleftrightarrow \quad \Delta\varepsilon_{ab} = 0 &&\quad\text{no interaction},\\
    &\eta_{ab} > 1 \quad \Longleftrightarrow \quad \Delta\varepsilon_{ab} < 0 &&\quad\text{activation}.
\end{align*}
We emphasize that an energy difference of $\infty$ is in fact a reasonable choice, considering one species might be able to fully block the other's binding site, leaving it no possibility to bind at all. Therefore, a state where both species are bound does not exist. Furthermore, we take a look at the case of $\Delta\varepsilon_{ab} = 0$. It can be seen from the probabilities that this indeed leaves us with no interaction as it reduces to
\begin{equation}
\label{eq: interaction_hill}
p_A = \frac{\eta_a a + \eta_a \eta_b ab}{1 +\eta_a a + \eta_b b + \eta_a \eta_b ab} = \frac{\eta_a a (1 + \eta_b b)}{(1 + \eta_a a)(1 + \eta_b b)} = \frac{\eta_a a }{1 + \eta_a a} = p_A^{hill}.
\end{equation}
Since the denominator of $p_A$ is by definition always larger than the numerator, we get monotonicity with respect to $\eta_{ab}$ (Fig. \ref{fig: probability interaction}), i.e.
\begin{equation}
\label{eq: monotonicity}
\frac{\eta_a a + \eta_a \eta_b \underline{\eta} ab}{1 +\eta_a a + \eta_b b + \eta_a \eta_b \underline{\eta} ab} < \frac{\eta_a a + \eta_a \eta_b \overline{\eta}ab}{1 +\eta_a a + \eta_b b + \eta_a \eta_b \overline{\eta} ab} \qquad \text{for} \quad \underline{\eta}<\overline{\eta}.
\end{equation}
Positive interactions lie above the Hill function, whereas negative interactions always remain below. Finally, we realize that following this derivation any kind of interaction of this type is in fact mutual. That means, if $A$ inhibits $B$, then $B$ also inhibits $A$. Likewise, if $A$ activates $B$, then $B$ also activates $A$.
Under the notion that transcriptional regulation occurs based on inhibition and auto-activation, previous models have proposed to use the product of two Hill functions \cite{Bessonnard2014, Tosenberger2017, Cang2021}. In the context of probabilities this would imply stochastic independence. In a system, in which $B$ inhibits $A$, the total probability of $A$ binding the product of two Hill functions largely underestimates the true binding probability even for $\eta_{ab}=0$:
\begin{equation}
    \label{eq: hill vs prob}
    p_A^{hill}(1 - p_B^{hill}) = \frac{\eta_a a}{1 + \eta_a a}\cdot \frac{1}{1 + \eta_b b} \leq \frac{\eta_a a}{1+ \eta_a a + \eta_b b} = p_A\big|_{\eta_{ab}=0}.
\end{equation}
Furthermore, the interpretation from reaction kinetics also often leads to introducing the sum of various Hill functions. However, this linkage can cause total "probabilities" greater than $1$. Consequently, this coupling might cause nonphysical behavior in the dynamical system like e.g. concentrations exceeding the possible maximum.

\subsection{Application to NANOG/GATA6/FGF}

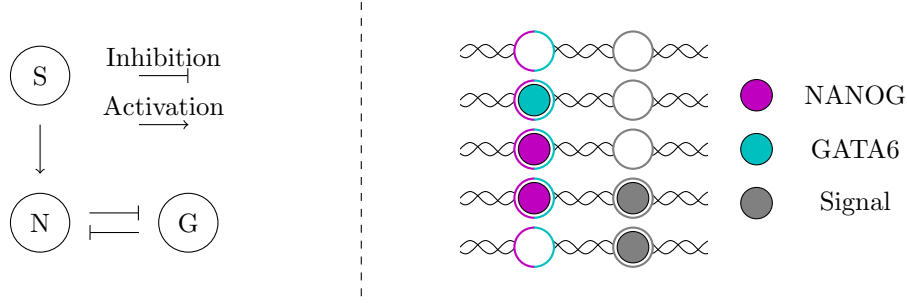
\begin{figure}
    \centering
    \begin{tikzpicture}[scale=1.3]
\newcommand\R{0.2}
\definecolor{NANOG}{rgb}{0.75,0,0.75}
\definecolor{GATA6}{rgb}{0,0.75,0.75}

\foreach \y in {0,0.5,1,1.5,2}
{
\begin{scope}
\clip  (0.25,-0.25-\y) rectangle (2.75,0.25-\y);
\draw [decorate, decoration={coil, aspect=0, segment length=15}] (0.25,0-\y) -- (6,0-\y);
\draw [decorate, decoration={coil, aspect=0, segment length=15}] (0,0-\y) -- (6,0-\y);
\end{scope}

\fill[color=white] (1,0-\y) circle (0.2);
\fill[color=white] (2,0-\y) circle (0.2);

\begin{scope}
\clip (1-\R-0.1,-\R-0.1-\y) rectangle (1,\R+0.1-\y);
\draw[thick, color=NANOG] (1,0-\y) circle (\R);
\end{scope}

\begin{scope}
\clip (1,-\R-0.1-\y) rectangle (1+\R+0.1,\R+0.1-\y);
\draw[thick, color=GATA6] (1,0-\y) circle (\R);
\end{scope}

\draw[thick, color=gray] (2,0-\y) circle (\R);
}

\draw[fill=GATA6] (1,-0.5) circle (0.8*\R);

\draw[fill=NANOG] (1,-1) circle (0.8*\R);

\draw[fill=NANOG] (1,-1.5) circle (0.8*\R);
\draw[fill=gray] (2,-1.5) circle (0.8*\R);

\draw[fill=gray] (2,-2) circle (0.8*\R);

\node (a) at (3.25, -0.45) {};
\node (b) at (3.25, -1) {};
\node (c) at (3.25, -1.55) {};

\draw[fill=NANOG] (a) circle (0.8*\R);
\draw[fill=GATA6] (b) circle (0.8*\R);
\draw[fill=gray] (c) circle (0.8*\R);

\node[align=left] at ($(a)+(1,0)$) {NANOG};
\node[align=left] at ($(b)+(1,0)$) {GATA6};
\node[align=left] at ($(c)+(1,0)$) {Signal};

\node (N) at (-4, -1.75) {N};
\node (G) at (-2.5, -1.75) {G};
\node (S) at (-4, -0.25) {S};
\draw (N) circle (1.5*\R);
\draw (G) circle (1.5*\R);
\draw (S) circle (1.5*\R);
\draw[-|] ($(N)+(0.5,0.1)$) -- ($(G)+(-0.5,0.1)$);
\draw[|-] ($(N)+(0.5,-0.1)$) -- ($(G)+(-0.5,-0.1)$);
\draw[->] ($(S)+(-0.0,-0.5)$) -- ($(N)+(0.0,0.5)$);
\draw[-|] (-3,-0.25) -- (-2.5,-0.25) node[midway, above] {Inhibition};
\draw[->] (-3,-0.75) -- (-2.5,-0.75) node[midway, above] {Activation};

\draw[style=dashed] (-0.75,0.5) -- (-0.75,-2.5);

\end{tikzpicture}
    \caption{Schematic representation of the GRN (left). NANOG (N) and GATA6 (G) mutually inhibit each other. The signal (S) activates N. The GRN is translated into a picture of the possible combinations of bindings to the DNA (right). Binding sites are shown as circles. NANOG and GATA6 share their binding site (bicolor border), whereas the signal gets a separate binding site (grey border). Bound states for each species are indicated by disks with the respective color. The five different cases represent all possible binding arrangements allowed in our model.}
    \label{fig: bindings}
\end{figure}

Previous work has provided a detailed insight into transcriptional regulation during early embryogenesis through a complicated model \cite{Bessonnard2014,Tosenberger2017}. We aim to reduce the complexity of the GRN by condensing it to the three essential building blocks: Normalized concentrations of NANOG $n$, GATA6 $g$ and the signal $s$. We interpret the signal, as the influence of neighboring cells acting on the cell via the Fgf/Erk signaling pathway. In contrast to the general consideration of modeling based on reaction kinetics, our approach is based only on the different possibilities for transcription factor binding (Fig. \ref{fig: bindings}). Therefore, we extract from the GRN one of the ways in which the different combinations of binding can look. Building our binding probabilities, we assume the inhibition to be of the blocking type, i.e. NANOG and GATA6 cannot simultaneously bind to a binding site. Consequently, a triple bound state, i.e. for which NANOG, GATA6 and the signal are bound simultaneously also does not exist. The signal is allowed to bind together with NANOG but not with GATA6. Hence, we chose interaction coefficients
\begin{equation}
    \label{eq: interaction coefficients}
    \eta_{ng} = \eta_{gs} = \eta_{ngs} = 0, \qquad \eta_{ns} \geq 1 \Longleftrightarrow -\Delta\varepsilon_{ns} > 0.
\end{equation}
Any single bound state results in the terms $\eta_\alpha \alpha$ with $\alpha \in \{n,g,s\}$. The remaining state has $n$ and $s$ bound simultaneously, yielding the term $\eta_n\eta_s\eta_{ns}ns$. For the binding probability of NANOG, we collect all the terms including $n$ and divide them by the combination of all other terms, resulting in
\begin{equation}
\label{p_NANOG}
    p_N = \frac{\eta_n n(1+\eta_s \eta_{ns} s)}{1 + \eta_n n(1+\eta_s \eta_{ns} s) + \eta_g g + \eta_s s}.
\end{equation}
Likewise, the probability of GATA6 is given by
\begin{equation}
\label{p_GATA6}
     p_G = \frac{\eta_g g}{1 + \eta_n n(1+\eta_s \eta_{ns} s) + \eta_g g + \eta_s s}.
\end{equation}

\subsection{Transcriptional ODE}

Our interest lies in the temporal evolution of the NANOG and GATA6 expressions. The transcription-translation process enables us to formulate the evolution of transcription factors via the binding probabilities. If $n$ is bound, it will be reproduced with a reproduction rate $r_n$. Simultaneously, $n$ decays over time with constant decay rate $\gamma_n$. Analogously, we apply this logic to $g$. Considering up to $M$ individual cells $i=1,...,M$ interacting with each other, this results in the following system of ordinary differential equations (ODEs):
\begin{equation}
\label{eq: ODE system}
\begin{aligned}
    \frac{dn}{dt} &= r_n \frac{\eta_n n_i(1+\eta_s \eta_{ns} s_i)}{1 + \eta_n n_i(1+\eta_s \eta_{ns} s_i) + \eta_g g_i + \eta_s s_i} - \gamma_n n_i \\
    \frac{dg}{dt} &= r_g \frac{\eta_g g_i}{1 + \eta_n n_i(1+\eta_s \eta_{ns} s_i) + \eta_g g_i + \eta_s s_i} - \gamma_g g_i, \qquad i = 1,...,M.
\end{aligned}
\end{equation}
We note that the cell-cell interactions are fully encoded in the signal $s_i$ obtained from every individual cell. In the course of this study, we restrict ourselves to a signal only depending on the GATA6 expressions of other cells, i.e. 
\begin{equation}
    \label{eq: generalized signal}
    \boldsymbol{s}: \mathbb{R}^M \to \mathbb{R}^M: \boldsymbol{g} \mapsto \boldsymbol{s}(\boldsymbol{g}), \qquad \frac{\partial s_i}{\partial g_i} = 0, \qquad i = 1,...,M.
\end{equation}
The condition on the right of \eqref{eq: generalized signal} guarantees that in this setting, the signal does not depend on the GATA6 expression of itself.

\section{Steady State Analysis}

\subsection{Steady states}

In order to get a better understanding of our ODE system, we want to delve further into the resulting steady states of the system. This means, we consider
$$\frac{dn_i}{dt} = 0 = \frac{dg_i}{dt}.$$
Consequently, we get
\begin{align}
\label{eq: steady_state_n}
    \frac{\eta_n n_i(1+\eta_s\eta_{ns}s_i)}{1 + \eta_n n_i(1+\eta_s\eta_{ns}s_i) + \eta_g g_i + \eta_s s_i} &= \frac{\gamma_n}{r_n} n_i, \\
\label{eq: steady_state_g}
    \frac{\eta_g g_i}{1 + \eta_n n_i(1+\eta_s\eta_{ns}s_i) + \eta_g g_i + \eta_s s_i} &= \frac{\gamma_g}{r_g} g_i.
\end{align}
When rearranging \eqref{eq: steady_state_n} and \eqref{eq: steady_state_g}, we find two possible solutions for $n_i$ and $g_i$, respectively. These solutions are
\begin{equation}
    \label{eq: solutions_n_g}
    n_i = \begin{cases}
    0 & \\
    \frac{r_n}{\gamma_n} - \frac{1+\eta_g g_i + \eta_s s_i}{\eta_n(1+\eta_s \eta_{ns} s_i)}
    \end{cases}, \qquad
    g_i = \begin{cases}
    0 & \\
    \frac{r_g}{\gamma_g} - \frac{1+\eta_n n_i(1+\eta_s \eta_{ns} s_i) + \eta_s s_i}{\eta_g}
    \end{cases}
\end{equation}
Taking every combination of $n_i$ and $g_i$ from \eqref{eq: solutions_n_g} into account, we end up with four different steady states.
For three of the steady states, we can get either no expression of NANOG and GATA6 or high expression of one transcription factor and none for the other:
\begin{alignat}{2}
    \label{eq: steady state 1}
    n_i &= 0, && \quad g_i = 0 \\
    \label{eq: steady state 2}
    n_i &= 0, && \quad g_i = \frac{r_g}{\gamma_g}-\frac{1+\eta_s s_i}{\eta_g} \\
    \label{eq: steady state 3}
    n_i &= \frac{r_n}{\gamma_n}-\frac{1+\eta_s s_i}{\eta_n(1+\eta_s \eta_{ns} s_i)}, && \quad g_i = 0 
\end{alignat}
These steady states share the lower bound $0$. Additionally, a rough estimate for an upper bound is given by the ratios of reproduction and decay $r_n/\gamma_n$ and $r_g/\gamma_g$. For parameter combinations such that 
\begin{equation}
     \frac{r_n}{\gamma_n} \gg \frac{1}{\eta_n}, \quad \frac{r_g}{\gamma_g} \gg \frac{1}{\eta_g} + \frac{\eta_s}{\eta_g}s_i
\end{equation}
the left hand sides of the inequalities provide a reliable estimate for the steady state values.\\
The fourth steady state is an oddity that arises by combining the non-zero solutions for $n_i$ and $g_i$ from \eqref{eq: solutions_n_g}. When combined, the corresponding variables $n_i$ and $g_i$ cancel out and we find the relation
\begin{equation}
    \label{eq: steady state 4 condition}
    \eta_n (1 + \eta_s \eta_{ns} s_i) = \eta_g \frac{r_g \gamma_n}{r_n \gamma_g}.
\end{equation}
This also leaves our system to be over-determined and the values of $n_i$ and $g_i$ cannot further be identified. However, by using \eqref{eq: steady state 4 condition} in the steady state solution $n_i \neq 0$ in \eqref{eq: solutions_n_g}, we obtain the following state:
\begin{equation}
    \label{eq: steady state 4}
    \frac{r_g\gamma_n}{r_n\gamma_g} n_i + g_i = \frac{r_g}{\gamma_g} - \frac{1+\eta_s s_i}{\eta_g}.
\end{equation}
For cell fate specification, \eqref{eq: steady state 4} is not relevant. For the simulations, we choose parameter values such that condition \eqref{eq: steady state 4 condition} cannot be satisfied. However, since the relation also depends on the signal, no general expression for the parameters can be derived at this point. We will come back to this later, once we have defined a concrete realization of our signal. Altogether, we have successfully identified the relevant steady states \eqref{eq: steady state 1}-\eqref{eq: steady state 3} of our ODE system \eqref{eq: ODE system}.

\subsection{Linearization}
In the following sections, we investigate the steady states in further detail. We employ linear stability analysis to determine the parameter regime that allows us to find a desired steady state for the overall system. At the single cell level, we rule out \eqref{eq: steady state 1}, since it is not relevant to cell fate specification. At the tissue level, we distinguish between homogeneous and heterogeneous steady states. A homogeneous equilibrium state consists of cells of a single type only. This means that either all of the cells in the tissue are in state \eqref{eq: steady state 2} or all of them are in state \eqref{eq: steady state 3}. To best reproduce the situation in the embryo, we want a mixture of two cell types. Therefore, we aim at excluding the homogeneous steady states as well.

We follow the definition of linear stability for an ODE system
$$\frac{dx_i}{dt} = f(x), \qquad i = 1,...,M.$$

We say, an ODE system is linearly stable in $x^*$, if its linearization matrix $L^{ODE} = f'(x^*)$ has only eigenvalues with negative real part. Using the $M$-dimensional identity matrix $I_M$, we can write the linearization matrix of \eqref{eq: ODE system} as
\begin{equation}
 \label{eq: linearization matrix}
L^{ODE} = \begin{pmatrix}                                
r_n\frac{\partial p_N}{\partial n} - \gamma_n I_M & r_n\frac{\partial p_N}{\partial g} \\                
r_g\frac{\partial p_G}{\partial n} & r_g\frac{\partial p_G}{\partial g} - \gamma_g I_M                   
\end{pmatrix},
\end{equation}
where we define $\frac{\partial p_N}{\partial n} := \left(\frac{\partial p_N}{\partial n_j,}(n_i,g_i,s_i)\right)_{i,j=1,...,M}$. The other block matrices are defined analogously. Specifically, we obtain

\begin{align}
    \label{eq: dpN/dn}
    \frac{\partial}{\partial n_j}p_N(n_i,g_i,s_i) &= \begin{cases} \frac{\eta_n(1+\eta_s \eta_{ns} s_i)(1+\eta_g g_i + \eta_s s_i)}{(1 + \eta_n n_i(1+\eta_s \eta_{ns} s_i) + \eta_g g_i + \eta_s s_i)^2}, & \qquad \text{if } i = j \\
    0 & \qquad \text{if } i \neq j
    \end{cases} \\
    \label{eq: dpN/dg}
    \frac{\partial}{\partial g_j}p_N(n_i,g_i,s_i) &= \begin{cases} -\frac{\eta_n \eta_g n_i(1+\eta_s \eta_{ns} s_i)}{(1 + \eta_n n_i(1+\eta_s \eta_{ns} s_i) + \eta_g g_i + \eta_s s_i)^2}, & \qquad \text{if } i = j \\
    \frac{\eta_n \eta_s \eta_{ns} n_i(1+\eta_g g_i) - \eta_n \eta_s n_i}{(1 + \eta_n n_i(1+\eta_s \eta_{ns} s_i) + \eta_g g_i + \eta_s s_i)^2}\frac{\partial s_i}{\partial g_j} & \qquad \text{if } i \neq j
    \end{cases} \\
    \label{eq: dpG/dn}
    \frac{\partial}{\partial n_j}p_G(n_i,g_i,s_i) &= \begin{cases} -\frac{\eta_n \eta_g g_i(1+\eta_s \eta_{ns} s_i)}{(1 + \eta_n n_i(1+\eta_s \eta_{ns} s_i) + \eta_g g_i + \eta_s s_i)^2}, & \qquad \text{if } i = j \\
    0 & \qquad \text{if } i \neq j
    \end{cases} \\
    \label{eq: dpG/dg}
    \frac{\partial}{\partial g_j}p_G(n_i,g_i,s_i) &= \begin{cases} \frac{\eta_g(1+\eta_n n_i(1+\eta_s \eta_{ns} s_i) + \eta_s s_i)}{(1 + \eta_n n_i(1+\eta_s \eta_{ns} s_i) + \eta_g g_i + \eta_s s_i)^2}, & \qquad \text{if } i = j \\
    -\frac{\eta_s \eta_g g_i(\eta_n \eta_{ns} n_i + 1)}{(1 + \eta_n n_i(1+\eta_s \eta_{ns} s_i) + \eta_g g_i + \eta_s s_i)^2}\frac{\partial s_i}{\partial g_j} & \qquad \text{if } i \neq j
    \end{cases}
\end{align}

 Thus, both the first and the third block matrix in \eqref{eq: linearization matrix} are diagonal, which significantly reduces the upcoming efforts in the stability analysis. Finding the eigenvalues of $L^{ODE}$ seems quite brutal at first but remember that for a system to be linearly unstable, only a single eigenvalue needs to be larger than zero. We can use the computational rules of the determinant for block matrices to write the characteristic polynomial as
\begin{equation}
    \label{eq: characteristic polynomial}
    \begin{aligned}
    \chi(\lambda) =& \det(L^{ODE}-\lambda I_{2M}) \\
    =& \det\left(r_n \frac{\partial p_N}{\partial n}-(\gamma_n+\lambda) I_M\right) \\
    &\cdot\det\left(r_g\frac{\partial p_G}{\partial g} -(\gamma_g+\lambda) I_M - r_g\frac{\partial p_G}{\partial n}\left[r_n\frac{\partial p_N}{\partial n}-(\gamma_n+\lambda) I_M\right]^{-1}r_n\frac{\partial p_N}{\partial g}\right)
    \end{aligned}
\end{equation}
As usual, the eigenvalues are defined as the roots of the characteristic polynomial.

\subsection{Steady state \eqref{eq: steady state 1}}

In the following, we elaborate on how to exclude the first steady state \eqref{eq: steady state 1} as solution for our ODE system \eqref{eq: ODE system}. Without loss of generality, we assume $n_1 = 0 = g_1$. It suffices to focus on the first factor of the characteristic polynomial \eqref{eq: characteristic polynomial}. We find
$$\frac{\partial}{\partial n_1}p_N(0,0,s_1) = \eta_n \frac{1+\eta_s \eta_{ns} s_1}{1+\eta_s s_1}.$$
Due to its diagonal structure, we find the very first factor of the complete determinant to be
$$r_n \eta_n \frac{1+\eta_s \eta_{ns} s_1}{1+\eta_s s_1} - \gamma_n - \lambda \stackrel{!}{=} 0.$$
This translates to the eigenvalue $\lambda$ being
$$\lambda = r_n\eta_n \frac{1+\eta_s \eta_{ns} s_1}{1+\eta_s s_1} - \gamma_n.$$
Now $\lambda > 0$ yields
$$\eta_n > \frac{\gamma_n}{r_n} \frac{1+\eta_s s_1}{1+\eta_s \eta_{ns} s_1}.$$
Although the signal thus far has not been further specified, we propose a realistic physical representation by assuming $s_i \geq 0$. Furthermore, we consider an activation of $n$ by the signal $s$, i.e. $\eta_{ns}>1$ and therefore, inequality
\begin{equation}
    \label{eq: coefficient restriction 1}
    \eta_n > \frac{\gamma_n}{r_n}
\end{equation}
and consequently
\begin{equation}
    \label{eq: energy restriction 1}
    -\Delta\varepsilon_n > \ln\left(\frac{\gamma_n}{r_n}\right)
\end{equation}
provides a necessary condition for instability. The exclusion of this steady state strengthens our focus on \eqref{eq: steady state 2} and \eqref{eq: steady state 3}, which represent the two different cell types PrE and Epi, respectively.

\subsection{Homogeneous steady state \eqref{eq: steady state 2}}

With steady states \eqref{eq: steady state 2} and \eqref{eq: steady state 3}, we aim to find a parameter region for which we achieve a heterogeneous steady state, i.e. we get a tissue with a mixture of cells in the two states. To this end, we derive conditions for instability of the homogeneous steady state. We start with state \eqref{eq: steady state 2} and set $n_i = 0$ and $g_i = \frac{r_g}{\gamma_g}-\frac{1+\eta_s s_i}{\eta_g}$ for all $i$. Inserting these expressions into the derivatives \eqref{eq: dpN/dn}-\eqref{eq: dpG/dg} results in a simplification of $L^{ODE}$. Since \eqref{eq: dpN/dg} is zero for every $i$, due to \eqref{eq: characteristic polynomial} the relevant derivatives are:
\begin{align}
    \left(\frac{\partial p_N}{\partial n}\right)_{i,i} &= \frac{\gamma_g\eta_n}{r_g\eta_g}\left(1+\eta_s \eta_{ns}s_i\right), \\
    \left(\frac{\partial p_G}{\partial g}\right)_{i,i} &= \frac{\gamma_g^2}{r_g^2} \frac{1+\eta_s s_i}{\eta_g}, \\
    \left(\frac{\partial p_G}{\partial g}\right)_{i,j} &=\frac{\eta_s}{\eta_g}\left(\frac{\gamma_g}{r_g} - \frac{\gamma_g^2}{r_g^2}\frac{1+\eta_s s_i}{\eta_g}\right) \frac{\partial s_i}{\partial g_j}.
\end{align}
Altogether, using this result in \eqref{eq: characteristic polynomial} leaves us with the polynomial
\begin{align*}
    \chi(\lambda) &= \det\left(r_n\frac{\partial p_N}{\partial n}-(\gamma_n+\lambda)I_M\right)\det\left(r_g\frac{\partial p_G}{\partial g}-(\gamma_g+\lambda)I_M\right) \\
    &= \left[\prod_{i=1}^M \gamma_g\frac{r_n\eta_n}{r_g\eta_g}(1+\eta_s \eta_{ns} s_i) - \gamma_n - \lambda\right] \det\left(r_g\frac{\partial p_G}{\partial g}-(\gamma_g+\lambda)I_M\right).
\end{align*}
The first factor already determines the first $M$ eigenvalues. For instability, it is sufficient that only one of these is greater than zero. In other words, this results in the inequality
$$\gamma_g\frac{r_n\eta_n}{r_g\eta_g}(1+\eta_s \eta_{ns} s_i) > \gamma_n.$$
After appropriate rearranging, we obtain a sufficient condition for our parameters
\begin{equation}
\label{eq: coefficient restriction 2}
\eta_g < \eta_n \frac{r_n \gamma_g}{r_g \gamma_n}(1+ \eta_s\eta_{ns} \max_i s_i).
\end{equation}
At this point, the general case cannot be simplified further. Depending on the cell-cell interaction and therefore the incoming signal $s_i$, one can find an even more accurate description of this relation. Alternatively, we can formulate this condition in terms of energy differences as
\begin{equation}
\label{eq: energy restriction 2}
 -\Delta \varepsilon_g < -\Delta \varepsilon_n  + \ln \left(1+ e^{-\Delta\varepsilon_s-\Delta\varepsilon_{ns}} \max_i s_i\right) + \ln \left(\frac{r_n \gamma_g}{r_g \gamma_n}\right),
\end{equation}
which allows us to see the maximum allowed deviation of the difference between $\Delta\varepsilon_n$ and $\Delta\varepsilon_g$. Keep in mind that for this condition, we only relied on the first $M$ eigenvalues. In truth, this condition might be even more relaxed than what we derived.

\subsection{Homogeneous steady state \eqref{eq: steady state 3}}

We set $n_i = \frac{r_n}{\gamma_n}-\frac{1+\eta_s s_i}{\eta_n(1+\eta_s \eta_{ns} s_i)}$ and $g_i = 0$. Using the same approach as before, we can neglect the off-diagonal matrices, since \eqref{eq: dpG/dn} is zero for all $i$. According to \eqref{eq: characteristic polynomial}, the relevant derivatives are then
\begin{align*}
    \left(\frac{\partial p_N}{\partial n}\right)_{i,i} &= \frac{\gamma_n^2}{r_n^2}\frac{1+\eta_s s_i}{\eta_n(1+\eta_s\eta_{ns}s_i)}, \\
    \left(\frac{\partial p_G}{\partial g}\right)_{i,i} &=  \frac{\gamma_n}{r_n}\frac{\eta_g}{\eta_n}\frac{1}{1 + \eta_s \eta_{ns}s_i}.
\end{align*}
The characteristic polynomial then becomes
\begin{equation*}
    \chi(\lambda) = \prod_{i=1}^M\left[\frac{\gamma_n^2}{r_n}\frac{1+\eta_s s_i}{\eta_n(1+\eta_s\eta_{ns}s_i)} -\gamma_n -\lambda\right] \cdot\prod_{i=1}^M\left[r_g\frac{\gamma_n}{r_n}\frac{\eta_g}{\eta_n}\frac{1}{1 + \eta_s \eta_{ns}s_i}-\gamma_g-\lambda\right].
\end{equation*}
We exploit again the instability condition that any eigenvalue must be positive and find two different inequalities
\begin{align}
    \label{eq: coefficient restriction 3.1}
    \eta_n & < \frac{\gamma_n}{r_n}\frac{1+\eta_s s_i}{1+\eta_s\eta_{ns}s_i}, \\
    \label{eq: coefficient restriction 3.2}
    \eta_g & > \frac{r_n \gamma_g}{r_g \gamma_n}\eta_n(1 + \eta_s \eta_{ns}s_i).
\end{align}
We remark, that condition \eqref{eq: coefficient restriction 3.1} lies in conflict with \eqref{eq: coefficient restriction 1} due to $\eta_{ns} > 1$ and is therefore neglected. Nevertheless, \eqref{eq: coefficient restriction 3.2} yields a condition for $\eta_g$. As before, it is necessary to fulfill this inequality for a single value $s_i$, i.e. the minimum of all possible signal values suffices in that regard
\begin{equation}
    \label{eq: coefficient restriction 3}
    \eta_g > \frac{r_n \gamma_g}{r_g \gamma_n}\eta_n(1 + \eta_s \eta_{ns}\min_i s_i).
\end{equation}
Again, we write this in terms of energy differences
\begin{equation}
    \label{eq: energy restriction 3}
    -\Delta\varepsilon_g  > -\Delta\varepsilon_n + \ln\left(1+ e^{-\Delta\varepsilon_s-\Delta\varepsilon_{ns}} \min_i s_i\right) + \ln \left(\frac{r_n \gamma_g}{r_g \gamma_n}\right).
\end{equation}

\subsection{Steady state summary}

The stability conditions \eqref{eq: energy restriction 2} and \eqref{eq: energy restriction 3} define an interval for $-\Delta\varepsilon_g$, 
\begin{equation}
    \label{eq: stability interval}
    \Delta\varepsilon_{min} < -\Delta\varepsilon_g < \Delta\varepsilon_{max}
\end{equation}
with
\begin{align}
    \Delta\varepsilon_{min} &:= -\Delta\varepsilon_n + \ln\left(1+ e^{-\Delta\varepsilon_s-\Delta\varepsilon_{ns}} \min_i s_i\right) + \ln \left(\frac{r_n \gamma_g}{r_g \gamma_n}\right) \\
    \Delta\varepsilon_{max} &:= -\Delta\varepsilon_n + \ln\left(1+ e^{-\Delta\varepsilon_s-\Delta\varepsilon_{ns}} \max_i s_i\right) + \ln \left(\frac{r_n \gamma_g}{r_g \gamma_n}\right)
\end{align}
The reproduction rates $r_n, r_g$ and decay rates $\gamma_n, \gamma_g$ shift this interval by $\ln \left(\frac{r_n \gamma_g}{r_g \gamma_n}\right)$. The length of the interval is determined by the minimum and maximum signal values combined with the associated energy differences $-\Delta\varepsilon_s$ and $-\Delta\varepsilon_{ns}$. In practice, $s_i$ depends on $\boldsymbol{g}$, which in return depends on $s_i$ which requires us to solve an equation to exactly find $\max_i s_i$. In order to avoid this, it is possible to choose parameters such that the steady state \eqref{eq: steady state 2} admits to an approximate solution
\begin{equation}
\label{eq: g approximation}
    g_i = \frac{r_g}{\gamma_g} - \frac{1+\eta_s s_i}{\eta_g} \approx \frac{r_g}{\gamma_g}.
\end{equation}
Depending on the nature of the signal $s_i$, this can be used to define a simplified stability interval.

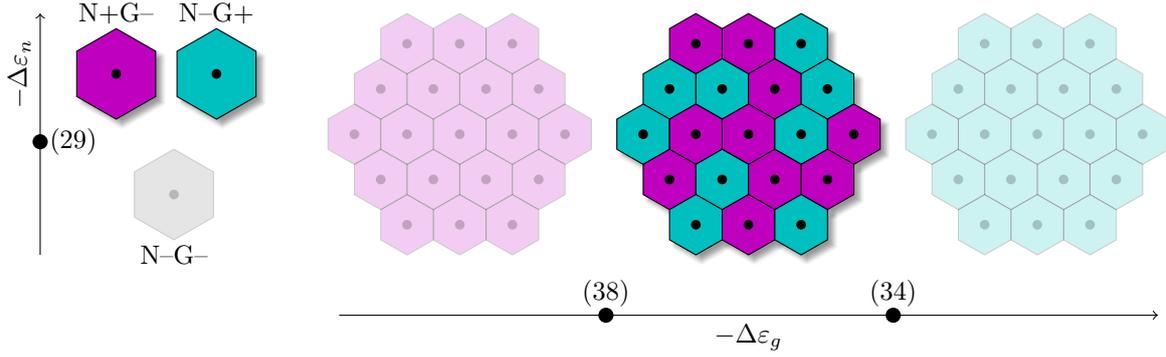
\begin{figure}
    \centering
    \begin{tikzpicture}

\definecolor{NANOG}{rgb}{0.75,0,0.75}
\definecolor{GATA6}{rgb}{0,0.75,0.75}

\newcommand\W{0.8}

\begin{scope}[%
every node/.style={anchor=west, regular polygon, 
regular polygon sides=6,
draw,
minimum width=1.5*\W cm,
outer sep=0,shape border rotate=90
},transform shape]
\node[fill=NANOG,blur shadow={shadow blur steps=5}] (b) at (-5*\W,-0.5*\W) {};
\fill (b) circle (0.08*\W);
\node[fill=gray, opacity=0.2] (a) at ($(b) + (0.3*\W,-2*\W)$) {};
\fill[opacity=0.2, fill opacity=0.2] (a) circle (0.08*\W);
\node[fill=GATA6,blur shadow={shadow blur steps=5}] (c) at ($(b) + (1*\W,0)$) {};
\fill (c) circle (0.08*\W);
\end{scope}

\node[align=left] at ($(a)+(0, -1*\W)$) {N--G--};
\node[align=left] at ($(b)+(0,1*\W)$) {N+G--};
\node[align=left] at ($(c)+(0,1*\W)$) {N--G+};

\fill ($(b) + (-1.25*\W,-1.125*\W)$) circle (0.1) node[right] {\eqref{eq: energy restriction 1}};
\draw[->] ($(b) + (-1.25*\W,-3*\W)$) -- ($(b) + (-1.25*\W,1*\W)$) node[near end, above, rotate=90] {$-\Delta\varepsilon_n$};

\begin{scope}[%
every node/.style={anchor=west, regular polygon, 
regular polygon sides=6,
draw,
minimum width=\W cm,
outer sep=0,shape border rotate=90
},
      transform shape]
    \node[fill=NANOG, opacity=0.2] (A1) {}; \fill[opacity=0.2, fill opacity=0.2] (A1) circle (0.08*\W);
    \node[fill=NANOG, opacity=0.2] (B1) at ($(A1)+(0.433*\W,0)$) {}; \fill[opacity=0.2, fill opacity=0.2] (B1) circle (0.08*\W);
    \node[fill=NANOG, opacity=0.2] (C1) at ($(B1)+(0.433*\W,0)$) {}; \fill[opacity=0.2, fill opacity=0.2] (C1) circle (0.08*\W);
    
    \node[fill=NANOG, opacity=0.2] (A2) at ($(A1)-(2*0.433*\W,3/4*\W)$) {}; \fill[opacity=0.2, fill opacity=0.2] (A2) circle (0.08*\W);
    \node[fill=NANOG, opacity=0.2] (B2) at ($(A2)+(0.433*\W,0)$) {}; \fill[opacity=0.2, fill opacity=0.2] (B2) circle (0.08*\W);
    \node[fill=NANOG, opacity=0.2] (C2) at ($(B2)+(0.433*\W,0)$) {}; \fill[opacity=0.2, fill opacity=0.2] (C2) circle (0.08*\W);
    \node[fill=NANOG, opacity=0.2] (D2) at ($(C2)+(0.433*\W,0)$) {}; \fill[opacity=0.2, fill opacity=0.2] (D2) circle (0.08*\W);
    
    \node[fill=NANOG, opacity=0.2] (A3) at ($(A2)-(2*0.433*\W,3/4*\W)$) {}; \fill[opacity=0.2, fill opacity=0.2] (A3) circle (0.08*\W);
    \node[fill=NANOG, opacity=0.2] (B3) at ($(A3)+(0.433*\W,0)$) {}; \fill[opacity=0.2, fill opacity=0.2] (B3) circle (0.08*\W);
    \node[fill=NANOG, opacity=0.2] (C3) at ($(B3)+(0.433*\W,0)$) {}; \fill[opacity=0.2, fill opacity=0.2] (C3) circle (0.08*\W);
    \node[fill=NANOG, opacity=0.2] (D3) at ($(C3)+(0.433*\W,0)$) {}; \fill[opacity=0.2, fill opacity=0.2] (D3) circle (0.08*\W);
    \node[fill=NANOG, opacity=0.2] (E3) at ($(D3)+(0.433*\W,0)$) {}; \fill[opacity=0.2, fill opacity=0.2] (E3) circle (0.08*\W);
    
    \node[fill=NANOG, opacity=0.2] (A4) at ($(A3)-(0,3/4*\W)$) {}; \fill[opacity=0.2, fill opacity=0.2] (A4) circle (0.08*\W);
    \node[fill=NANOG, opacity=0.2] (B4) at ($(A4)+(0.433*\W,0)$) {}; \fill[opacity=0.2, fill opacity=0.2] (B4) circle (0.08*\W);
    \node[fill=NANOG, opacity=0.2] (C4) at ($(B4)+(0.433*\W,0)$) {}; \fill[opacity=0.2, fill opacity=0.2] (C4) circle (0.08*\W);
    \node[fill=NANOG, opacity=0.2] (D4) at ($(C4)+(0.433*\W,0)$) {}; \fill[opacity=0.2, fill opacity=0.2] (D4) circle (0.08*\W);
    
    \node[fill=NANOG, opacity=0.2] (A5) at ($(A4)-(0,3/4*\W)$) {}; \fill[opacity=0.2, fill opacity=0.2] (A5) circle (0.08*\W);
    \node[fill=NANOG, opacity=0.2] (B5) at ($(A5)+(0.433*\W,0)$) {}; \fill[opacity=0.2, fill opacity=0.2] (B5) circle (0.08*\W);
    \node[fill=NANOG, opacity=0.2] (C5) at ($(B5)+(0.433*\W,0)$) {}; \fill[opacity=0.2, fill opacity=0.2] (C5) circle (0.08*\W);
\end{scope}

\begin{scope}[%
every node/.style={anchor=west, regular polygon, 
regular polygon sides=6,
draw,
minimum width=\W cm,
outer sep=0,shape border rotate=90
},
      transform shape]
    \node[fill=GATA6, opacity=0.2] (A1) at (9.5*\W, 0){}; \fill[opacity=0.2, fill opacity=0.2] (A1) circle (0.08*\W);
    \node[fill=GATA6, opacity=0.2] (B1) at ($(A1)+(0.433*\W,0)$) {}; \fill[opacity=0.2, fill opacity=0.2] (B1) circle (0.08*\W);
    \node[fill=GATA6, opacity=0.2] (C1) at ($(B1)+(0.433*\W,0)$) {}; \fill[opacity=0.2, fill opacity=0.2] (C1) circle (0.08*\W);
    
    \node[fill=GATA6, opacity=0.2] (A2) at ($(A1)-(2*0.433*\W,3/4*\W)$) {}; \fill[opacity=0.2, fill opacity=0.2] (A2) circle (0.08*\W);
    \node[fill=GATA6, opacity=0.2] (B2) at ($(A2)+(0.433*\W,0)$) {}; \fill[opacity=0.2, fill opacity=0.2] (B2) circle (0.08*\W);
    \node[fill=GATA6, opacity=0.2] (C2) at ($(B2)+(0.433*\W,0)$) {}; \fill[opacity=0.2, fill opacity=0.2] (C2) circle (0.08*\W);
    \node[fill=GATA6, opacity=0.2] (D2) at ($(C2)+(0.433*\W,0)$) {}; \fill[opacity=0.2, fill opacity=0.2] (D2) circle (0.08*\W);
    
    \node[fill=GATA6, opacity=0.2] (A3) at ($(A2)-(2*0.433*\W,3/4*\W)$) {}; \fill[opacity=0.2, fill opacity=0.2] (A3) circle (0.08*\W);
    \node[fill=GATA6, opacity=0.2] (B3) at ($(A3)+(0.433*\W,0)$) {}; \fill[opacity=0.2, fill opacity=0.2] (B3) circle (0.08*\W);
    \node[fill=GATA6, opacity=0.2] (C3) at ($(B3)+(0.433*\W,0)$) {}; \fill[opacity=0.2, fill opacity=0.2] (C3) circle (0.08*\W);
    \node[fill=GATA6, opacity=0.2] (D3) at ($(C3)+(0.433*\W,0)$) {}; \fill[opacity=0.2, fill opacity=0.2] (D3) circle (0.08*\W);
    \node[fill=GATA6, opacity=0.2] (E3) at ($(D3)+(0.433*\W,0)$) {}; \fill[opacity=0.2, fill opacity=0.2] (E3) circle (0.08*\W);
    
    \node[fill=GATA6, opacity=0.2] (A4) at ($(A3)-(0,3/4*\W)$) {}; \fill[opacity=0.2, fill opacity=0.2] (A4) circle (0.08*\W);
    \node[fill=GATA6, opacity=0.2] (B4) at ($(A4)+(0.433*\W,0)$) {}; \fill[opacity=0.2, fill opacity=0.2] (B4) circle (0.08*\W);
    \node[fill=GATA6, opacity=0.2] (C4) at ($(B4)+(0.433*\W,0)$) {}; \fill[opacity=0.2, fill opacity=0.2] (C4) circle (0.08*\W);
    \node[fill=GATA6, opacity=0.2] (D4) at ($(C4)+(0.433*\W,0)$) {}; \fill[opacity=0.2, fill opacity=0.2] (D4) circle (0.08*\W);
    
    \node[fill=GATA6, opacity=0.2] (A5) at ($(A4)-(0,3/4*\W)$) {}; \fill[opacity=0.2, fill opacity=0.2] (A5) circle (0.08*\W);
    \node[fill=GATA6, opacity=0.2] (B5) at ($(A5)+(0.433*\W,0)$) {}; \fill[opacity=0.2, fill opacity=0.2] (B5) circle (0.08*\W);
    \node[fill=GATA6, opacity=0.2] (C5) at ($(B5)+(0.433*\W,0)$) {}; \fill[opacity=0.2, fill opacity=0.2] (C5) circle (0.08*\W);
\end{scope}

\begin{scope}[%
every node/.style={anchor=west, regular polygon, 
regular polygon sides=6,
draw,
minimum width=\W cm,
outer sep=0,shape border rotate=90,blur shadow={shadow blur steps=5}
},
      transform shape]
    \node[fill=NANOG] (A1) at (4.75*\W,0) {}; \fill (A1) circle (0.08*\W);
    \node[fill=NANOG] (B1) at ($(A1)+(0.433*\W,0)$) {}; \fill (B1) circle (0.08*\W);
    \node[fill=GATA6] (C1) at ($(B1)+(0.433*\W,0)$) {}; \fill (C1) circle (0.08*\W);
    
    \node[fill=GATA6] (A2) at ($(A1)-(2*0.433*\W,3/4*\W)$) {}; \fill (A2) circle (0.08*\W);
    \node[fill=GATA6] (B2) at ($(A2)+(0.433*\W,0)$) {}; \fill (B2) circle (0.08*\W);
    \node[fill=NANOG] (C2) at ($(B2)+(0.433*\W,0)$) {}; \fill (C2) circle (0.08*\W);
    \node[fill=GATA6] (D2) at ($(C2)+(0.433*\W,0)$) {}; \fill (D2) circle (0.08*\W);
    
    \node[fill=GATA6] (A3) at ($(A2)-(2*0.433*\W,3/4*\W)$) {}; \fill (A3) circle (0.08*\W);
    \node[fill=NANOG] (B3) at ($(A3)+(0.433*\W,0)$) {}; \fill (B3) circle (0.08*\W);
    \node[fill=NANOG] (C3) at ($(B3)+(0.433*\W,0)$) {}; \fill (C3) circle (0.08*\W);
    \node[fill=GATA6] (D3) at ($(C3)+(0.433*\W,0)$) {}; \fill (D3) circle (0.08*\W);
    \node[fill=NANOG] (E3) at ($(D3)+(0.433*\W,0)$) {}; \fill (E3) circle (0.08*\W);
    
    \node[fill=NANOG] (A4) at ($(A3)-(0,3/4*\W)$) {}; \fill (A4) circle (0.08*\W);
    \node[fill=GATA6] (B4) at ($(A4)+(0.433*\W,0)$) {}; \fill (B4) circle (0.08*\W);
    \node[fill=NANOG] (C4) at ($(B4)+(0.433*\W,0)$) {}; \fill (C4) circle (0.08*\W);
    \node[fill=NANOG] (D4) at ($(C4)+(0.433*\W,0)$) {}; \fill (D4) circle (0.08*\W);
    
    \node[fill=GATA6] (A5) at ($(A4)-(0,3/4*\W)$) {}; \fill (A5) circle (0.08*\W);
    \node[fill=NANOG] (B5) at ($(A5)+(0.433*\W,0)$) {}; \fill (B5) circle (0.08*\W);
    \node[fill=GATA6] (C5) at ($(B5)+(0.433*\W,0)$) {}; \fill (C5) circle (0.08*\W);
\end{scope}

\fill ($(A3) - (0.6125*\W,3*\W)$) circle (0.1) node[above] {\eqref{eq: energy restriction 3}};
\fill ($(A3) - (-3.5*\W - 0.6125*\W,3*\W)$) circle (0.1) node[above] {\eqref{eq: energy restriction 2}};
\draw[->] ($(A3) - (5*\W,3*\W)$) -- ($(E3) - (-5*\W,3*\W)$) node[midway, below] {$-\Delta\varepsilon_g$};

\end{tikzpicture}
    \caption{Illustration of the different steady states at the single cell level (left) and the tissue level (right). The states we are aiming for are highlighted with higher opacity. Nodes and their corresponding number on the axes reference the relevant equation for the transition from one state to another.}
    \label{fig: steady states}
\end{figure}

The results of our stability analysis are summarized in figure \ref{fig: steady states}. From here on, we will designate the cells with high NANOG expression and low GATA6 expression from steady state \eqref{eq: steady state 3} as N+G--. Analogously, \eqref{eq: steady state 1} describes N--G-- cells and \eqref{eq: steady state 2} N--G+ cells.  At the single cell level, we are able to exclude N--G-- cells using inequality \eqref{eq: energy restriction 1}. Therefore, at the tissue level, we can distinguish between three different states. The stability interval \eqref{eq: stability interval} yields the exact parameter regime for the transition of the homogeneous states to the heterogeneous ones. These elegant lower and upper bounds for $-\Delta\varepsilon_g$ incorporate every parameter in our ODE system \eqref{eq: ODE system}. Finally, we know that the lower bound in \eqref{eq: stability interval} is associated with the homogeneous N--G+ state, whereas the upper bound is associated with the homogeneous N+G-- state. Therefore, we expect a monotonous increase in the number of N--G+ cells as the energy difference $-\Delta\varepsilon_g$ increases.

\section{Simulations}

\subsection{Cell arrangement}

\begin{figure}[htbp]
\centering
\includegraphics[width=0.49\textwidth]{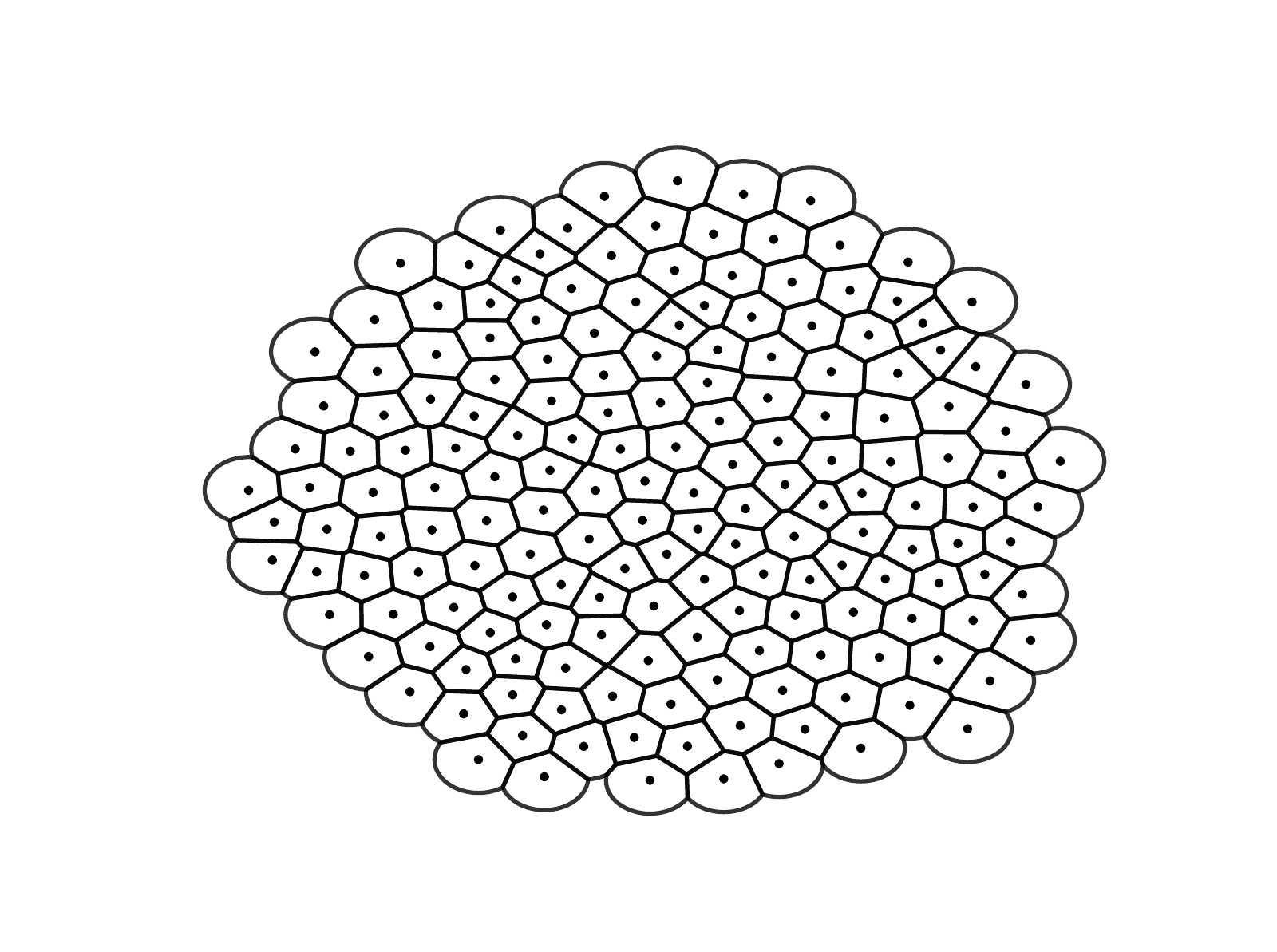}
\caption{Visualization of cells in a two-dimensional tissue. Points represent the center of mass of the cell nucleus. Lines show the respective cell boundaries. The tissue was generated according to \cite{Middleton2014, Stichel2017} starting with $9$ cells. The resulting tissue has 177 cell.}
\label{fig: tissue}
\end{figure}

For our following simulations, we use a two-dimensional representation of a cell tissue with 177 cells that was created based on an existing tissue-growth model \cite{Middleton2014, Stichel2017, Liebisch2020} (Fig. \ref{fig: tissue}). The number of cells was chosen as a point in time of the tissue growth and carries no further significance. Without shifting our focus too much on this part, we highlight the relevant steps of the tissue growth simulations:
\begin{itemize}
    \item Initialization of a fixed number of cells with position and radius close to each other
    \item Logistic growth of the radius
    \item Probabilistic cell division based on the radius
    \item Adhesion and repulsion result from overdamped motion using the Morse potential
\end{itemize}
This approach is capable of generating two-dimensional tissues. It is not limited to two spatial dimensions and can in fact be used for any positive dimension. For visualization purposes, we used a cutoff Voronoi tessellation to better highlight the size of the cell (Fig. \ref{fig: tissue}). The evolution of the transcription factors is carried out at the points for the cell nuclei, whereas shared cell boundaries are used to determine the neighborhood relations.

\subsection{Direct neighbor signaling}

\begin{figure}
    \centering
    \begin{tikzpicture}
\begin{scope}[%
every node/.style={anchor=west, regular polygon, 
regular polygon sides=6,
draw,
minimum width=2.5cm,
outer sep=0,
},
      transform shape]
    \node (A) {};
    \node (B) at (A.corner 1) {};
\end{scope}

\node (N1) at ($(A) + (-0.5,0)$) {N};
\node (G1) at ($(A) + (0.5,0)$) {G};
\draw[|-] ($(G1)+(-0.25,0.1)$) -- ($(N1)+(0.25,0.1)$);
\draw[-|] ($(G1)-(0.25,0.1)$) -- ($(N1)-(-0.25,0.1)$);

\node (N2) at ($(B) + (-0.5,0)$) {N};
\node (G2) at ($(B) + (0.5,0)$) {G};
\draw[|-] ($(G2)+(-0.25,0.1)$) -- ($(N2)+(0.25,0.1)$);
\draw[-|] ($(G2)-(0.25,0.1)$) -- ($(N2)-(-0.25,0.1)$);

\draw[->] (G1) edge[bend right] (N2);
\draw[->] (G2) edge[bend right=60] (N1);
\end{tikzpicture}
    \caption{Cell-cell interactions shown for two adjacent cells. NANOG and GATA6 again display their mutual inhibition. The signal is incorporated into a direct activation going from one cell to the other. Here, GATA6 activates NANOG in the neighboring cell.}
    \label{fig: GRN}
\end{figure}
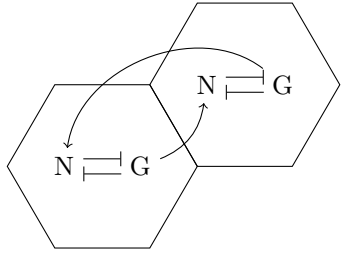

The complicated nature of signaling between ICM cells makes this an exciting field of research. We use a generic description of the signaling that aligns perfectly with our previously defined model. We propose the signaling protein concentration $\boldsymbol{s}$ to directly depend on GATA6 concentrations $\boldsymbol{g}$. Either via slow diffusion and a comparably fast uptake of these signals or specifically designed pathways to neighboring cells, we can imagine a process that allows cells to only interact with their direct neighbors. We then write the signal as
\begin{equation}
 \label{eq: signal}
    s_i = \frac{1}{|N_G(i)|}\sum_{j \in N_G(i)} g_j,\\
\end{equation}
where we used the notation $N_G(i)$ from graph theory to denote the neighbors of vertex $i$ in the Delaunay graph $G$. Here, we used the average number of neighbors as a weight to the signal. This approach is closely related to the one in \cite{Bessonnard2014, Tosenberger2017, Stanoev2021}. However, since we do not include the detailed dynamics of the signal, the receiving signal of a cell must not depend on the GATA6 expression levels of itself. We can also write the whole signal in terms of an adjacency matrix
\begin{equation}
    \label{eq: adjacency matrix}
    A = (A_{i,j})_{i,j=1,...,M}, \qquad A_{i,j} = \begin{cases}
    \frac{1}{N_G(i)} & \text{if } j \in N_G(i)\\
    0 & \text{if } j \notin N_G(i)
    \end{cases} \\
\end{equation}
such that $\boldsymbol{s} = A\boldsymbol{g}$. Using this definition, the cell-cell interaction can be described as an activation of NANOG through the GATA6 concentrations in the neighboring cells (Fig. \ref{fig: GRN}).

\subsection{Pattern formation}
\begin{figure}[htbp]
\centering
\begin{subfigure}{.32\textwidth}
  \centering
  \includegraphics[width=\linewidth]{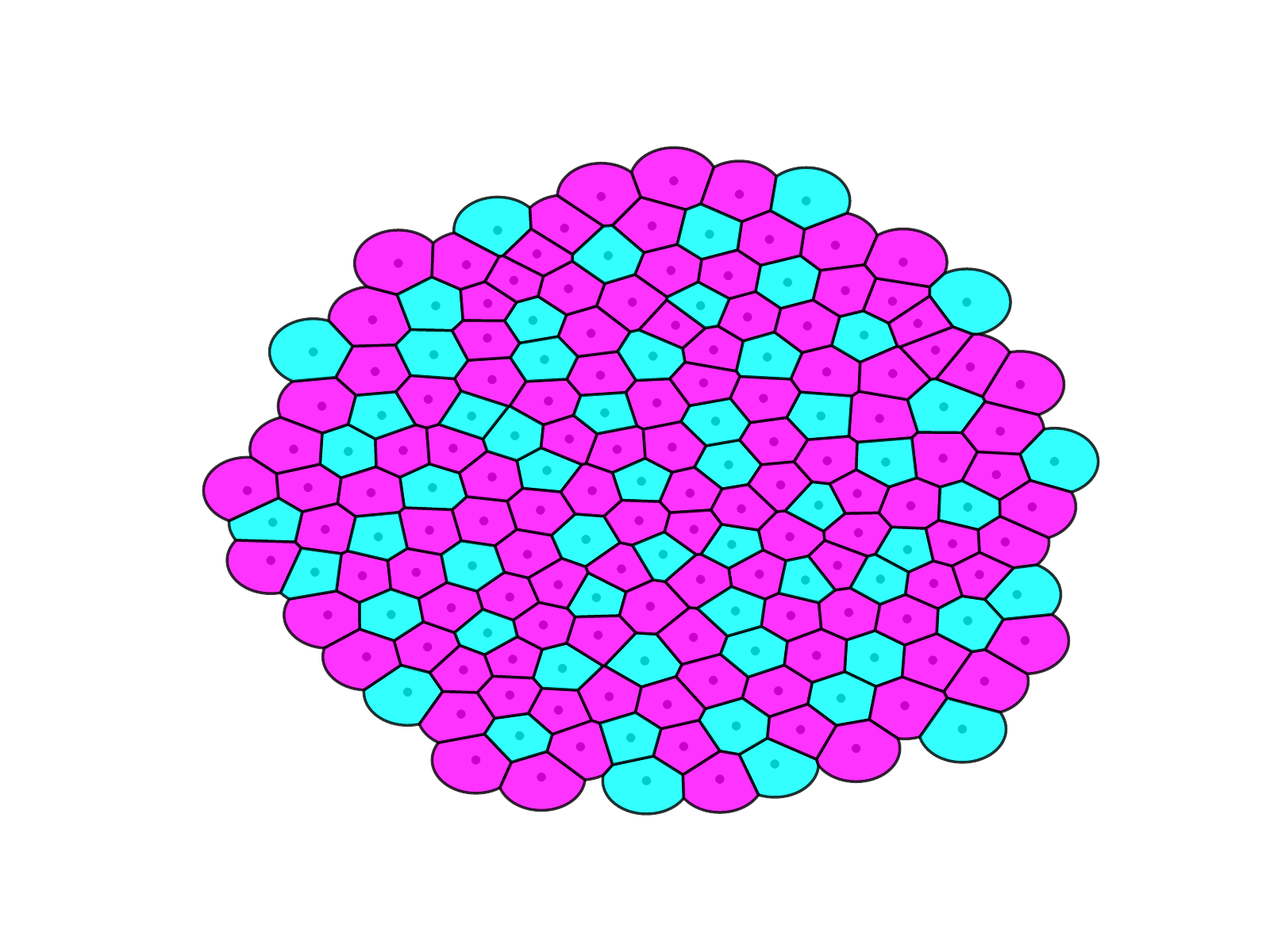}
  \subcaption{$\begin{aligned}
      -\Delta\varepsilon_g &= 7 \\
      N:G &= 2:1
  \end{aligned}$}
\end{subfigure}
\begin{subfigure}{.32\textwidth}
  \centering
  \includegraphics[width=\linewidth]{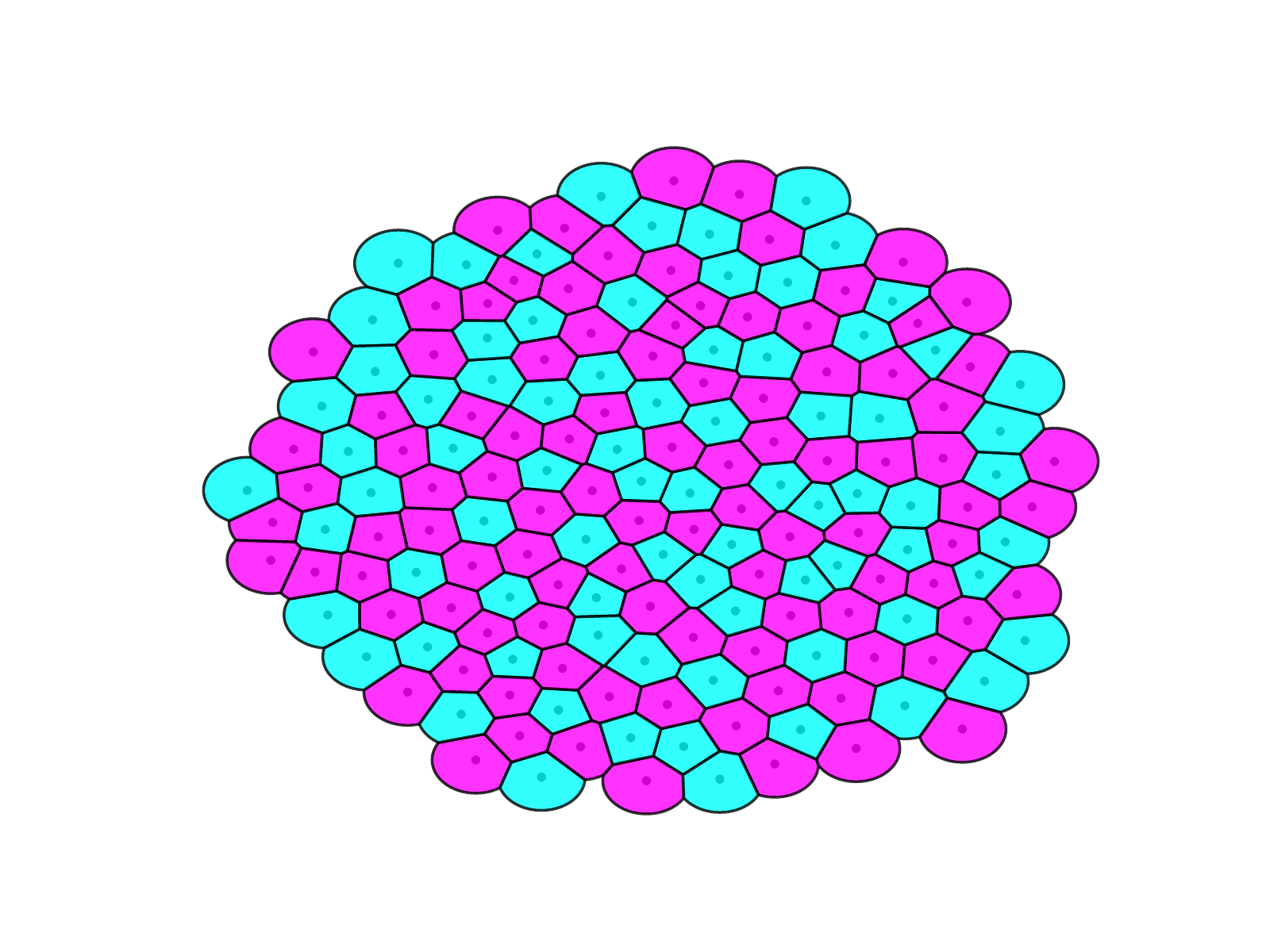}
  \subcaption{$\begin{aligned}
      -\Delta\varepsilon_g &= 7.31 \\
      N:G &= 99:78
  \end{aligned}$}
\end{subfigure}
\begin{subfigure}{.32\textwidth}
  \centering
  \includegraphics[width=\linewidth]{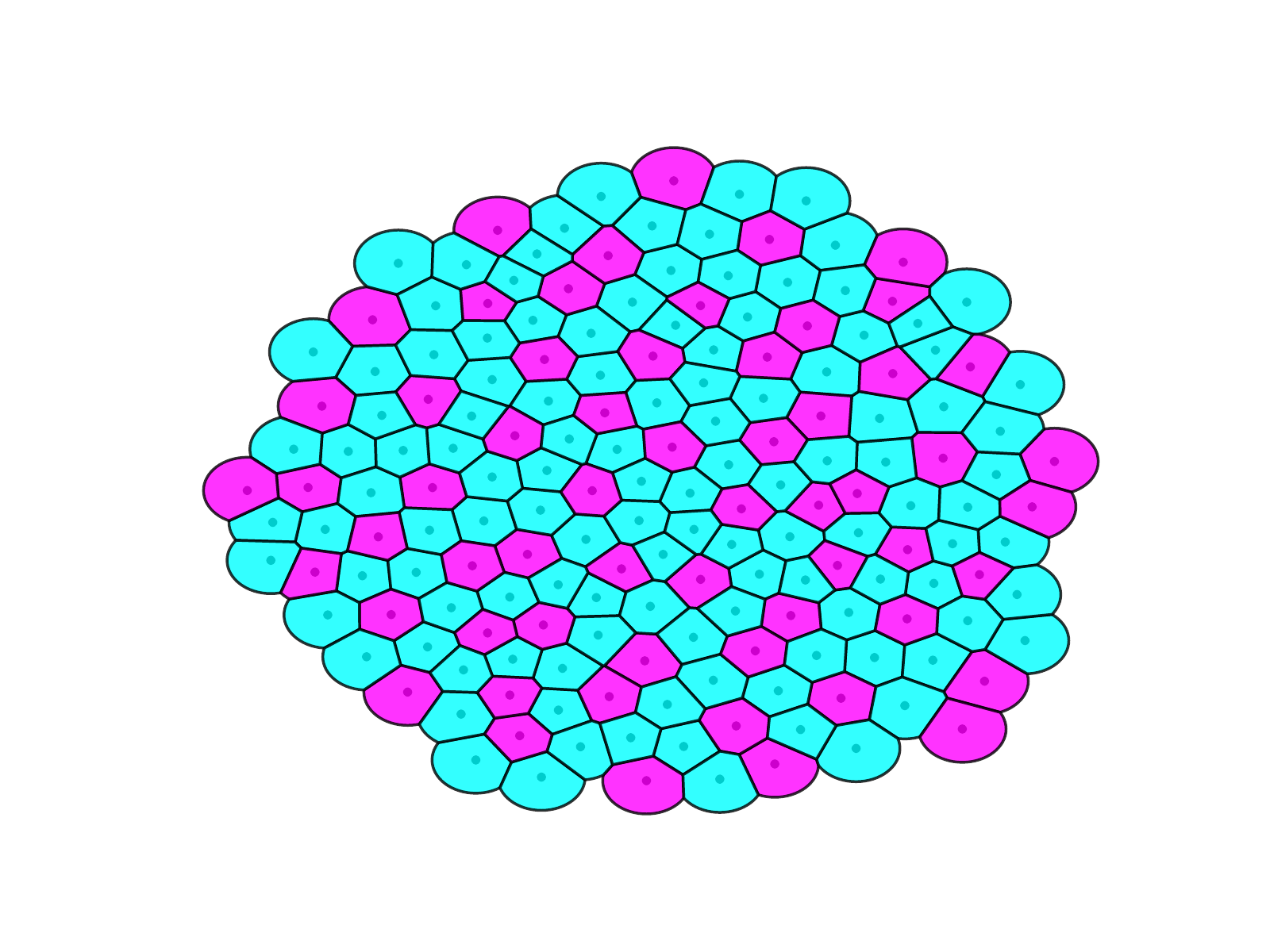}
  \subcaption{$\begin{aligned}
      -\Delta\varepsilon_g &= 7.6 \\
      N:G &= 1:2
  \end{aligned}$}
\end{subfigure}
\caption{Checkerboard pattern for three different ratios of N+G-- and N--G+ cells ($N:G$). The coloring uses the cell's NANOG expression levels. High NANOG expressions are colored in magenta, low NANOG expressions (high GATA6) in cyan.}
\label{fig: salt_and_pepper}
\end{figure}

Models of cell differentiation characterized by lateral inhibition tend to form an approximate checkerboard pattern of cells \cite{Collier1996} with a trend towards alternating cell types wherever possible. Our goal in this section is to show that our model is also capable of forming checkerboard patterns through lateral activation. The parameters used in any of the following simulations are fixed to $-\Delta\varepsilon_n = 6$, $-\Delta\varepsilon_n =-\Delta\varepsilon_{ns} = 2$, $r_n = r_g = 1$ and $\gamma_n = \gamma_g = 10$. The remaining energy difference $-\Delta\varepsilon_g$ is varied based on \eqref{eq: stability interval} to influence the cell type ratio. In the resulting cell fate pattern, N+G-- cells mostly avoid other N+G-- cells in their neighborhood (Fig. \ref{fig: salt_and_pepper}). The same behavior is also observed for N--G+ cells.

\subsection{Cell type proportions}
ICM organoids show a wide variety of different cell type proportions \cite{Mathew2019}. Furthermore, the ratio of N+G-- cells to N--G+ cells in wild type embryos is precise and likely crucial for further embryonic development \cite{Saiz2016, Saiz2020}. We analyze the range of possible cell type proportions and their dependence on our parameters. The signal \eqref{eq: signal} is linear in $\boldsymbol{g}$ and the lower bound for the signal is given by $0$. Using the normalization and a rough upper bound yields $s_i \leq \max_i g_i < r_g/\gamma_g = 0.1$. For the chosen parameter values, we get coefficients $\eta_s = 7.39$ and $\eta_n = 403.43$. We dismantle the terms in steady state \eqref{eq: steady state 2} to find
\begin{equation}
    \frac{1}{\eta_g} + \frac{\eta_s s_i}{\eta_g} \leq 0.0025 + 0.0018 \ll 0.1 = \frac{r_n}{\gamma_g}.
\end{equation}
Hence, approximation \eqref{eq: g approximation} is valid and leads to $\max_i s_i \approx r_g/\gamma_g$. Thus, \eqref{eq: stability interval} simplifies to
\begin{equation}
 -\Delta\varepsilon_n < -\Delta\varepsilon_g < -\Delta\varepsilon_n + \ln\left(1+ \eta_s\eta_{ns} \frac{r_g}{\gamma_g}\right).
\end{equation}
In our simulations, this yields the following bounding intervals
\begin{equation}
\label{eq: bounding interval simulation}
    \eta_g \in (403.43, 2606.08) \qquad \Longleftrightarrow \qquad -\Delta\varepsilon_g \in (6,7.87).
\end{equation}

\begin{figure}[htbp]
\centering
\includegraphics[width=0.49\textwidth]{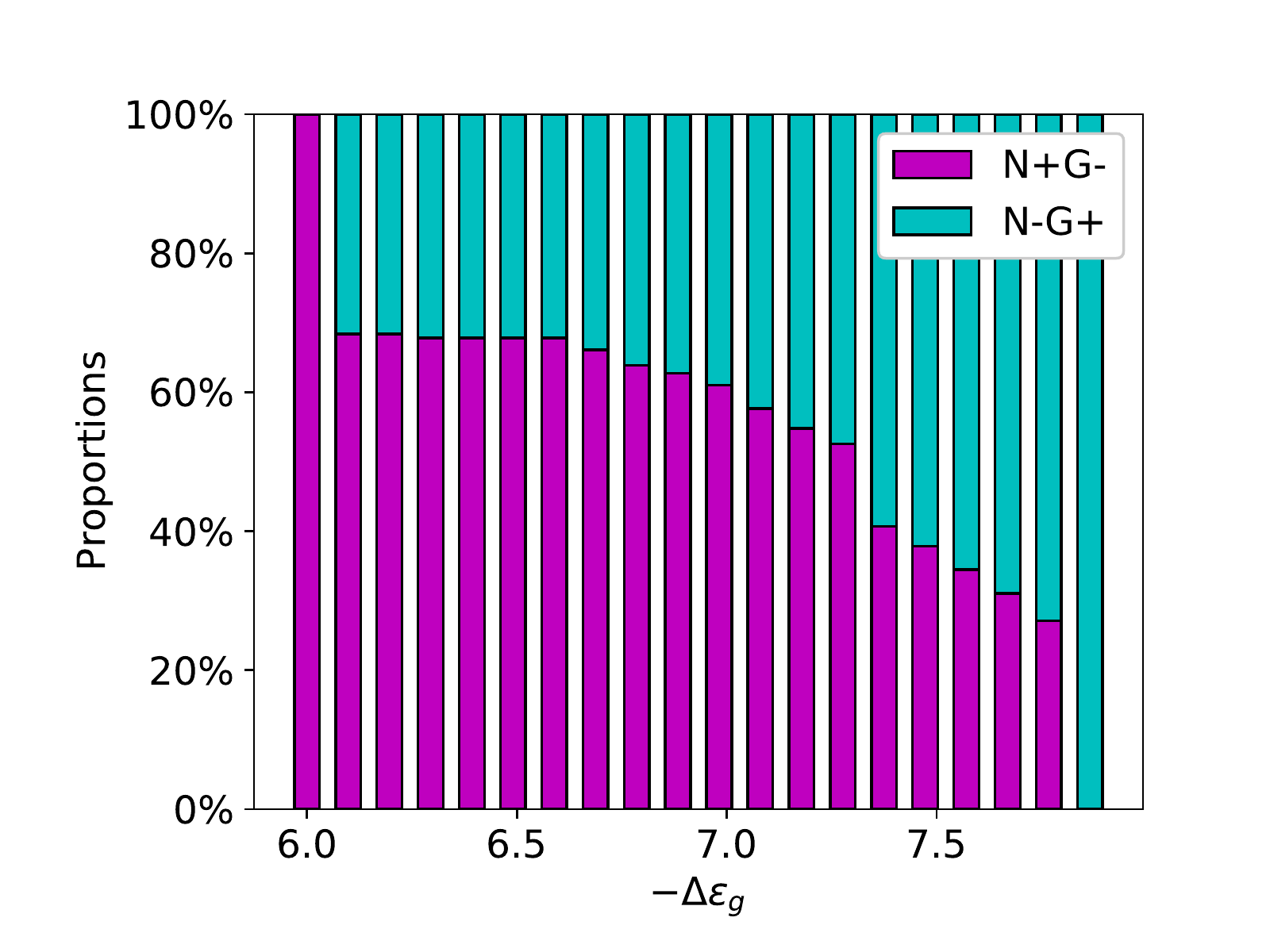}
\caption{Simulated cell type proportions for $20$ equidistant values of $-\Delta\varepsilon_g$ spanning over the stability interval \eqref{eq: stability interval}. N+G-- cell proportions are colored in magenta, N--G+ in cyan.}
\label{fig: proportions_local}
\end{figure}

The various cell type proportions (Fig. \ref{fig: proportions_local}) were simulated dividing the bounding interval \eqref{eq: bounding interval simulation} into $20$ equidistant values for $-\Delta\varepsilon_g$. The simulation results underline the result of the stability analysis. At the left and right boundaries we achieve homogeneity. In between, increasing $-\Delta\varepsilon_g$ yields a monotonous transition from only N+G-- to only N--G+ cells. The boundary regions suggest that proportions with about $70\%$ of one cell type and $30\%$ of the other are the maximum and minimum cell proportions achievable before reaching homogeneity. We hypothesize that these jumps are a result of the irregularity of the geometry itself. By this, we mean the different number of neighbors for every cell. In an idealized geometry these sharp transitions seem to vanish (see Appendix).

\subsection{Cell number}
\begin{figure}[htbp]
\centering
\begin{subfigure}{.32\textwidth}
  \centering
  \includegraphics[width=\linewidth]{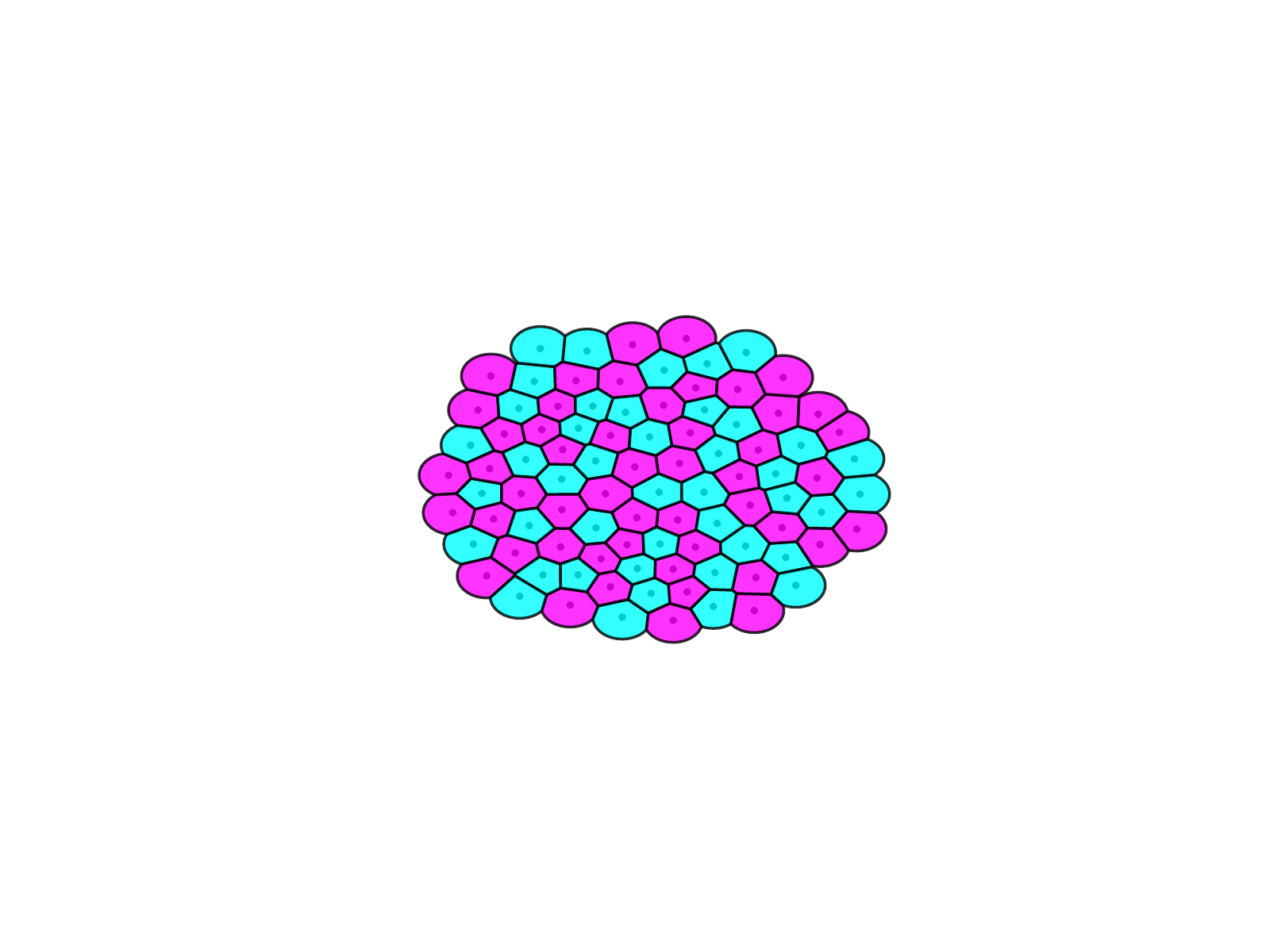}
\end{subfigure}
\begin{subfigure}{.32\textwidth}
  \centering
  \includegraphics[width=\linewidth]{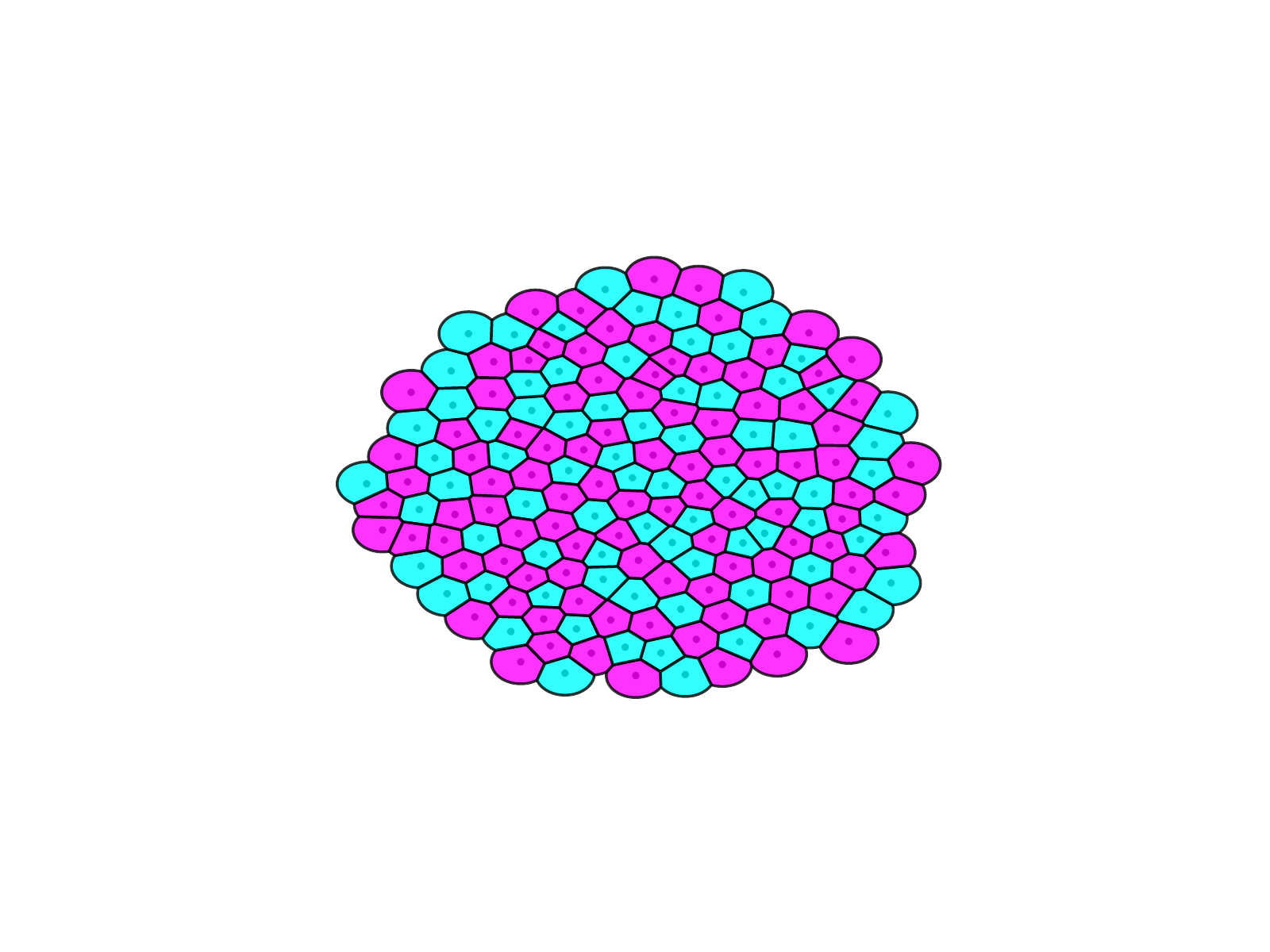}
\end{subfigure}
\begin{subfigure}{.32\textwidth}
  \centering
  \includegraphics[width=\linewidth]{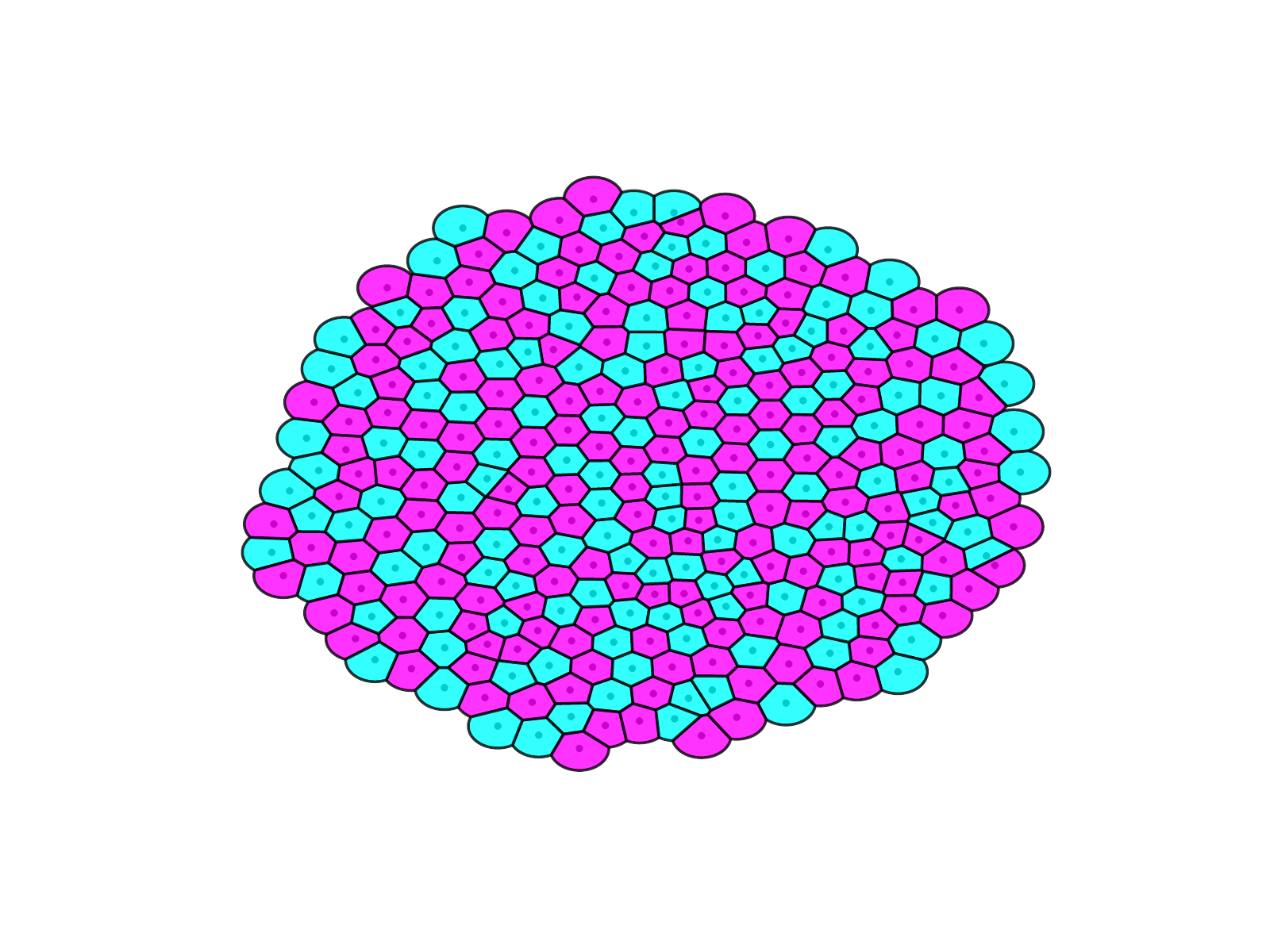}
\end{subfigure}
\caption{Final pattern for tissues of $93$ (left), $177$ (middle), and $324$ cells (right). Simulations use the same set of parameter values as before and $-\Delta\varepsilon_g = 7.31$. Cells with high NANOG expression are colored in magenta, cells with low NANOG expression in cyan. For comparison, we visualize the tissues on the same spatial scale.}
\label{fig: local size comparison}
\end{figure}

\begin{figure}[htbp]
\centering
  \includegraphics[width=0.49\linewidth]{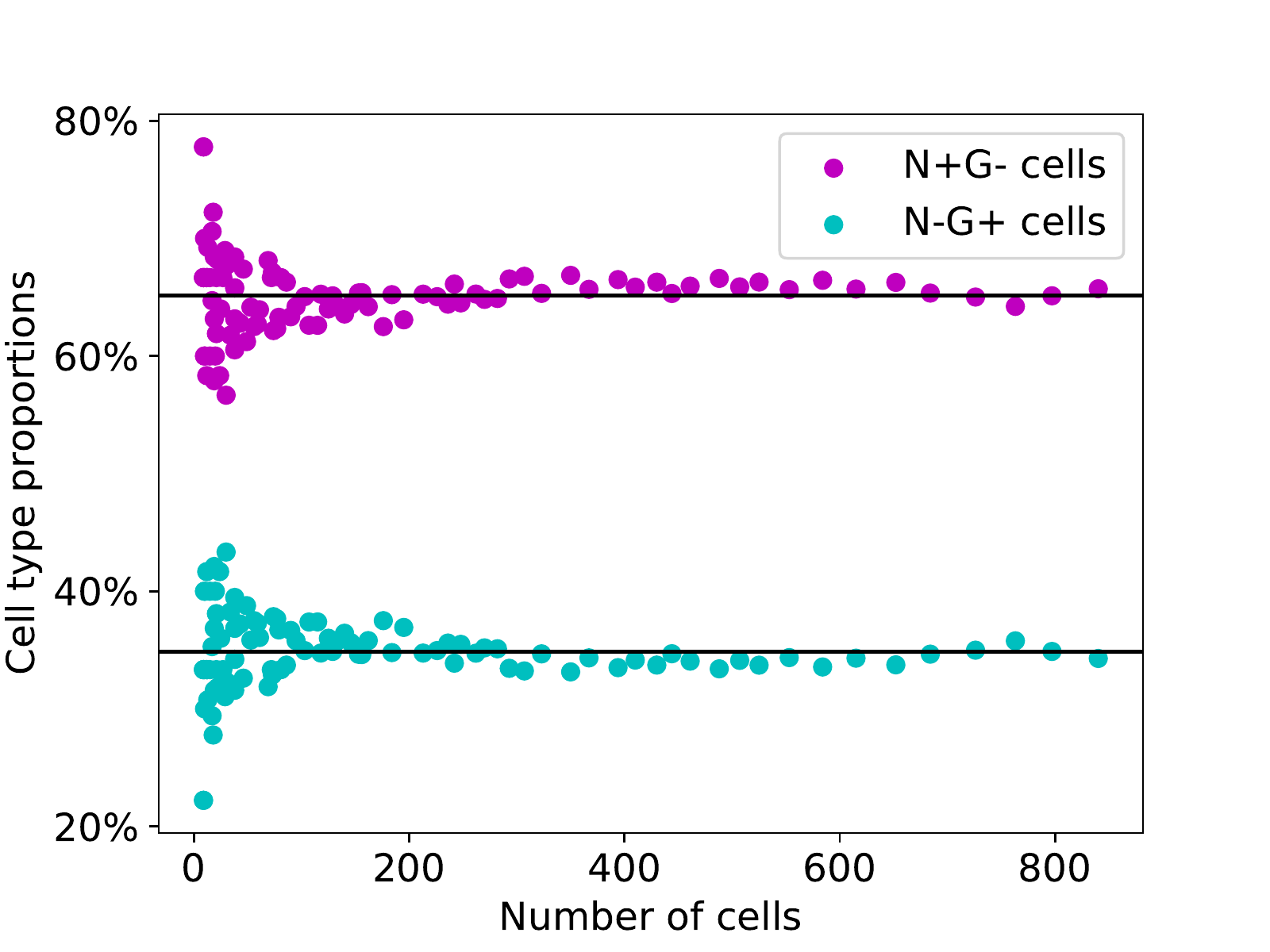}
\caption{Cell type proportions versus $100$ different cell numbers $M$. The cell numbers arise from stopping the tissue generation algorithm at equidistant points in time. Since the number of cells grows exponentially in time, the cell numbers are exponentially distributed. Simulations were carried out for $-\Delta\varepsilon_g = 7$. Vertical black lines depict the mean values of the respective cell type.}
\label{fig: poportions wrt size}
\end{figure}

ICM organoids come with different cell numbers while still often showing similar cell fate patterns \cite{Mathew2019}. To test whether the cell number has an effect on the overall pattern in our model, we considered two additional model geometries with $93$ and $324$ cells. The simulations used the same parameter values as above and $-\Delta\varepsilon_g = 7.31$. The checkerboard pattern is  robust with respect to the number of cells in the tissue (Fig. \ref{fig: local size comparison}). This is in line with our expectation, since cells in our model are only influenced by direct neighbors. The ratios of N+G-- to N--G+ cells are $1.16$, $1.27$ and $1.33$ for increasing cell number. Hence, we observe a minimal increase. To analyse this in more detail, we generated $100$ model geometries with different cell numbers to run our simulations on. We find that the cell type proportions remain approximately constant with respect to the number of cells (Fig. \ref{fig: poportions wrt size}). The discrete nature of the system together with deviations from a perfect circular geometry influence the number of neighbors for each cell. This effect particularly prevails with low cell numbers giving rise to fluctuations with a standard deviation of approximately $3.22~\%$. Together, both the checkerboard pattern and the cell type proportions are robust to changes in the number of cells.

\subsection{Discussion} 
Statistical mechanics has already proven its usefulness in biological model systems like ion channel opening and closing as well as oxygen hemoglobin binding \cite{Garcia2011}. These ideas have further been investigated for transcriptional regulation and were successfully applied for a wide variety of examples \cite{Bintu2005_1, Bintu2005_2}. To our knowledge, cell fate decision models have not been combined with statistical mechanics to date. We derived a generalized model directly from physical principles describing the cell fate specification of PrE versus Epi cells in preimplantation mouse embryos. Previously, the exclusivity of NANOG and GATA6 in late stages of development, and hence the existence of two different cell types, has been interpreted as the result of mutual inhibition of these two transcription factors within a cell \cite{Bessonnard2014, Tosenberger2017}. Spatial patterns of two different cell types further require intercellular interactions \cite{Collier1996}. Starting from a generalized signal, we subsequently describe cell-cell communication by a lateral activation from GATA6 in a cell to NANOG in the neighboring cells. 

We developed our model by applying a statistical mechanics approach \cite{Garcia2011}. The model distances itself from previous modeling approaches \cite{Bessonnard2014, Tosenberger2017} by a restriction to the few sufficient building blocks of cell differentiation in the inner cell mass, i.e. NANOG, GATA6 and a signal that allows cells to communicate. At the same time, it differs from other reduced models for intercellular signaling \cite{Stanoev2021} by using specifically derived binding probabilities instead of various applications of the Hill function. Counting the number of possible microstates for transcriptional regulation, we obtain the binding probabilities for NANOG and GATA6 that form the core of the model. Without any interactions between the transcription factors, our derivation results in the well-known Hill function. Extending this approach to interactions between NANOG and GATA6 of neighboring cells, results in a single binding probability for each transcription factor. A comparison with previous models highlights potential issues of the often used phenomenological models involving Hill functions \cite{Bessonnard2014, Tosenberger2017, Stanoev2021, Cang2021}. Linking multiple Hill functions by multiplication results in an underestimation of the true binding probabilities. In the context of probabilities the product would also require stochastic independence of the transcription factors. The sum of Hill functions can lead to binding "probabilities" larger than one and hence nonphysical behavior.

In the final model, we consider a tissue with a fixed number of cells. Within each cell, reproduction of transcription factors depending on their binding probabilities is balanced by exponential decay. A thorough steady state analysis including linear stability analysis, resulted in conditions for the two cell types of interest for mouse embryo development: high NANOG expression and low GATA6 expression (N+G--, Epi) or vice versa (N--G+, PrE). At the tissue level, linear stability analysis revealed additional parameter constraints that influence the decision between a homogeneous and a heterogeneous distribution of the cell types. Overall, this leads to a very high tractability of our model, which stands out from other models for mouse embryo development \cite{Bessonnard2014, Tosenberger2017}. 

To complement the stability analysis, we performed numerical simulations. Simulations were carried out on two-dimensional tissues inspired by the ICM organoids developed in \cite{Mathew2019}. We specified the intercellular signal as direct neighbor signal, such that GATA6 of one cell activates NANOG in the neighboring cell. This allowed the reproduction of the characteristic checkerboard pattern that has been observed in previous models \cite{Bessonnard2014, Tosenberger2017, Stanoev2021}. The the spatial pattern and the cell type proportions show robustness with respect to cell numbers. In particular, the latter is in good agreement with experimental observations on constant cell type proportions in mouse blastocysts with perturbed cell numbers \cite{Saiz2020}. Making use of the parameter restriction for the energy differences introduced the possibility to calibrate the model for various cell type proportions. This parameter constraint is characterized mainly by its dependence on the incoming signal, so that the signal plays an important role in terms of cell proportions. This result matches experimental findings for PrE differentiation in an in vitro stem cell culture \cite{Schroeter2015}. There, FGF signaling has been identified as the control for the proportion of PrE cells in the system.  

Using statistical mechanics to model transcriptional regulation in cells led to a very accessible and controllable ODE system for cell fate specification in preimplantation mouse embryos. It establishes a first link to our previous studies on mouse blastocysts and ICM organoids \cite{Mathew2019, Fischer2020} using a theoretical description of the pattern formation with two different cell types. The modeling is subject to a somewhat more challenging procedure, but ultimately leads to a system that is easier to tackle overall. Our modeling approach is deliberately general in nature to allow for an application to other GRNs. This facilitates the development of new and interesting models with improved physical interpretation.

\appendix
\beginAppendix

\section{Cell type proportions in ideal geometries}

\begin{figure}[htbp]
\centering
\includegraphics[width=0.49\textwidth]{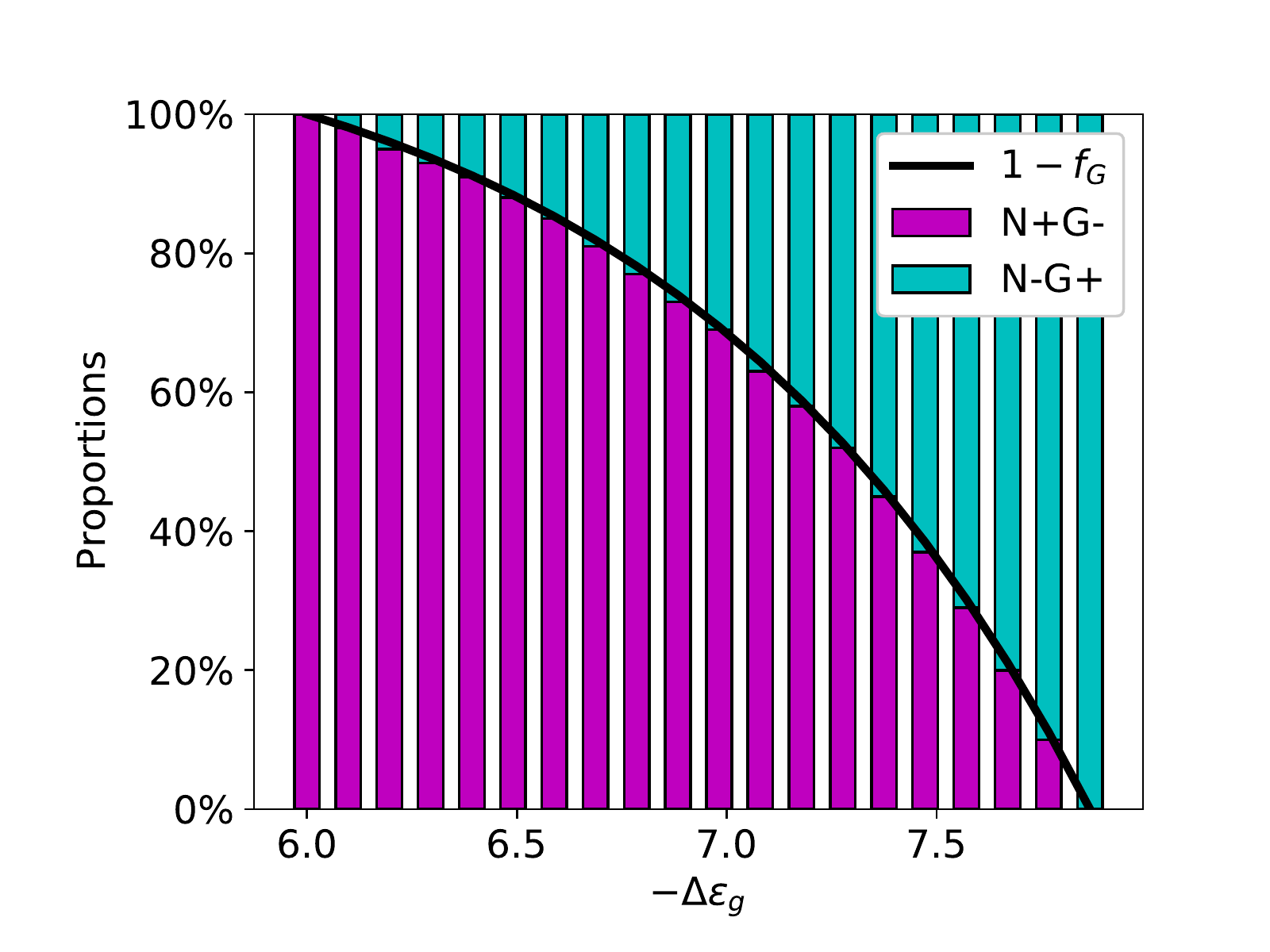}
\caption{Cell type proportions for a tissue with 10x10 square cells with periodic boundary conditions for $20$ equidistant values of $-\Delta\varepsilon_g$ within the stability interval \eqref{eq: stability interval}. N+G-- cell proportions are colored in magenta, N--G+ in cyan. The black curve marks our prediction for the cell type proportion according to \eqref{eq: proportions ideal energy}. Please note that the top bars represent the proportion of N+G-- cells and therefore, we plotted $1-f_G$.}
\label{fig: proportions ideal}
\end{figure}

We perform an analytical analysis of the relation of the cell type proportions and the parameter $-\Delta\varepsilon_g$. We define an ideal geometry such that the number of neighbors for each cell is equal. We explore a regular grid of $k \times k$ square cells. By introducing periodic boundaries, any cell in the system has exactly four neighbors. Pattern formation on similar grids has been investigated previously \cite{Collier1996}. Focusing on one cell in the tissue, we come back to the fourth steady state \eqref{eq: steady state 4}, i.e. non-zero solutions for both transcription factors, and its necessary condition \eqref{eq: steady state 4 condition} yields
\begin{equation}
\label{eq: tipping point}
    \eta_n (1 + \eta_s \eta_{ns} s_i) = \eta_g \frac{r_g \gamma_n}{r_n \gamma_g}.
\end{equation}
This equation describes the tipping point of a cell's fate. Both sides represent the numerator of a respective binding probability \eqref{p_NANOG} and \eqref{p_GATA6}. This means that deviating from "$=$" to "$>$" will increase the binding probability for NANOG, tipping its fate towards N+G--. Analogously, "$<$" will lead to N--G+. The signal $s_i$ depends only on neighboring cells and the cells themselves are all equal in terms of their neighborhood. Therefore, the signal becomes an approximate representation of the cell type proportions for ideal geometries. At first, we isolate $s_i$ in \eqref{eq: tipping point} to find
\begin{equation}
    s_i = \frac{r_g \gamma_n}{r_n \gamma_g} \frac{\eta_g - \eta_n}{\eta_n\eta_s \eta_{ns}}.
\end{equation}
By definition \eqref{eq: signal}, $s_i$ is the mean of a cells neighboring $g_j$ values. Assuming the neighbors to be in steady state and using the same steady state approximation as before, i.e.
\begin{equation}
    g_j = 0 \qquad \text{or} \qquad g_j \approx \frac{r_g}{\gamma_g}, \qquad j \in N_G(i), \quad i \in \{1, ..., M\}.
\end{equation}
the signal can be written as a fraction
\begin{equation}
    s_i = \frac{l}{4} \frac{r_g}{\gamma_g}, \qquad l \in \{0,1,2,3,4\}.
\end{equation}
A cell of N--G+ fate supports a maximum of $l^{\max}$ N--G+ cells in its neighborhood, where
\begin{equation}
\label{eq: l_max}
    l^{\max} := \left\lfloor 4\frac{\gamma_n}{r_n} \frac{\eta_g - \eta_n}{\eta_n\eta_s \eta_{ns}}\right\rfloor.
\end{equation}
Here, $\lfloor x \rfloor$ describes the floor function, i.e. the nearest lower integer of a number $x$. A single cell neighborhood can only mimic the true cell type proportions this far. However, in an ideal geometry with enough cells we hypothesize that many of these single cell neighborhoods organize themselves such that the prefactor $l/4$ can be replaced by a rational number that describes the total cell type proportions in the tissue. We therefore define the proportions of N--G+ cells as a function $\hat{f}_G$ of $\eta_g$ with
\begin{equation}
\label{eq: proportions ideal}
    \hat{f}_G(\eta_g) = \frac{\gamma_n}{r_n} \frac{\eta_g - \eta_n}{\eta_n\eta_s \eta_{ns}}.
\end{equation}
When formulating \eqref{eq: proportions ideal} in terms of energy differences, we get
\begin{equation}
\label{eq: proportions ideal energy}
    f_G(-\Delta\varepsilon_g) = \frac{\gamma_n}{r_n} \frac{e^{-\Delta\varepsilon_g} - e^{-\Delta\varepsilon_n}}{e^{-\Delta\varepsilon_n -\Delta\varepsilon_s -\Delta\varepsilon_{ns}}}.
\end{equation}
Simulation results show that the function $f_G$ provides an accurate representation of how the cell type proportions can be determined in an ideal geometry (Fig. \ref{fig: proportions ideal}). 

\bibliography{mybib.bib}
\end{document}